%% file: RIS-Aided_Cell-Free.tex
\documentclass[10pt,journal,final,twocolumn,]{IEEEtran}
\IEEEoverridecommandlockouts

\usepackage{amsmath,amssymb,amsfonts}
\usepackage{cite}
\usepackage{graphicx}
\usepackage{epstopdf}

\usepackage{booktabs}
\usepackage{graphicx}
\usepackage{textcomp}
\usepackage{xcolor}
\usepackage{times}
\usepackage{subfigure}         
\usepackage{amssymb,amsmath}
\usepackage{acronym}  
\usepackage{balance}

\usepackage{bm}         
\usepackage{algorithm} 
\usepackage{algorithmic}

\usepackage{lipsum}
\usepackage{stfloats}

\usepackage{url}

\usepackage{amsmath}
\allowdisplaybreaks[4]

\newtheorem{proposition}{\bf Proposition}

\newtheorem{remark}{\bf Remark}

\acrodef{ofdm}[OFDM]{orthogonal frequency division multiplexing}%
\acrodef{miso-ofdm}[MISO-OFDM]{multi-input single-output orthogonal frequency division multiplexing}%

\acrodef{ris}[RIS]{reconfigurable intelligent surface}%
\acrodef{qos}[QoS]{quality of service}%
\acrodef{idft}[IDFT]{inverse discrete Fourier transform}%
\acrodef{dft}[DFT]{discrete Fourier transform}%
\acrodef{cp}[CP]{cyclic prefix}%
\acrodef{csi}[CSI]{channel state information}%
\acrodef{awgn}[AWGN]{additive white Gaussion noise}%

\acrodef{qcqp}[QCQP]{quadratically constrained quadratic program}%
\acrodef{qp}[QP]{quadratic program}%
\acrodef{bs}[BS]{base station}%
\acrodef{ap}[BS]{base station}%
\acrodef{aps}[APs]{access points}%
\acrodef{qos}[QoS]{quality of service}%
\acrodef{ue}[UE]{user equipment}%
\acrodef{snr}[SNR]{signal-to-noise ratio}%
\acrodef{mmwave}[mmWave]{millimeter-wave}%
\acrodef{snr}[SNR]{signal-to-noise ratio}%

\acrodef{sinr}[SINR]{signal-to-interference-plus-noise ratio}%

\acrodef{ser}[SER]{symbol error rate}%
\acrodef{rc}[RC]{reflection coefficient}%
\acrodef{uavs}[UAVs]{unmanned aerial vehicles}%
\acrodef{mimo}[MIMO]{multiple-input multiple-output}%
\acrodef{noma}[NOMA]{non-orthogonal multiple access}%

\acrodef{ace}[ACE]{adaptive cross-entropy}%
\acrodef{wsr}[WSR]{weighted sum-rate}%
\acrodef{udn}[UDN]{ultra-dense network}%
\acrodef{Udn}[UDN]{Ultra-dense network}%

\def\BibTeX{{\rm B\kern-.05em{\sc i\kern-.025em b}\kern-.08em
    T\kern-.1667em\lower.7ex\hbox{E}\kern-.125emX}}

\begin{document}
\title{A Joint Precoding Framework for Wideband Reconfigurable Intelligent Surface-Aided \\Cell-Free Network}
\author{{{Zijian Zhang} and {Linglong Dai\vspace*{-2em}}
}

\thanks{This paper was presented in part at  the  IEEE SPAWC'20, Atlanta, GA, USA,  May 26–29,  2020 \cite{Zijian'20}.}

\thanks{All authors are with the Beijing National Research Center for Information Science and Technology (BNRist) as well as the Department of Electronic Engineering, Tsinghua University, Beijing 100084, China (e-mails: zhangzj20@mails.tsinghua.edu.cn, daill@tsinghua.edu.cn).}
\thanks{This work was supported in part by the National Key Research and Development Program of China (Grant No. 2020YFB1807201), in part by the National Natural Science Foundation of China (Grant No. 62031019), and in part by the European Commission through the H2020-MSCA-ITN META WIRELESS Research Project under Grant 956256. {\it(Corresponding author: Linglong Dai.)}}
}
\maketitle
\begin{abstract}
Thanks to the strong ability against the inter-cell interference, cell-free network is considered as a promising technique to improve network capacity. However, further capacity improvement requires to deploy more base stations (BSs) with high cost and power consumption. To address this issue, inspired by the recently developed reconfigurable intelligent surface (RIS) technique, we propose the concept of RIS-aided cell-free network to improve the capacity with low cost and power consumption. The key idea is to replace some of the required BSs by low-cost and energy-efficient RISs. Then, in a wideband RIS-aided cell-free network, we formulate the problem of joint precoding design at BSs and RISs to maximize the network capacity. Due to the non-convexity and high complexity of the formulated problem, we develop an alternating optimization framework to solve this challenging problem. In particular, we decouple this problem via fractional programming, and solve the subproblems alternatively. Note that most of the scenarios considered in existing works are special cases of the general scenario studied in this paper, and the proposed joint precoding framework can serve as a general solution to maximize the capacity in most existing RIS-aided scenarios. Finally, simulation results demonstrate that, compared with the conventional cell-free network, the network capacity under the proposed scheme can be improved significantly.
\end{abstract}
\section{Introduction}
\IEEEPARstart NETWORK technique is the most essential technique to increase the capacity of wireless communication systems \cite{D'15}. Compared with 4G, the capacity of 5G wireless network is expected to be increased by 1000 times \cite{Idachaba'16}. In the currently deployed cellular networks, all users in a cell are mainly served by one \ac{bs}, thus the users close to the cell boundary usually suffer from the severe inter-cell interference, which is caused by the signals from adjacent cells.
\par
\ac{Udn} has been proposed as a promising technique for 5G to further enhance the network capacity \cite{Kamel'16}. The core idea of UDN is to increase the number of \ac{bs}s and deploy small cells \cite{Andrews'16} in the \textit{cell-centric} cellular network. However, as the cell density increases, the inter-cell interference grows larger and larger. The cooperation theory has determined that \cite{Lozano'13}, the upper limit of the network capacity will be bounded by the inter-cell interference as long as the cell-centric network is used. In other words, inter-cell interference becomes the bottleneck for the capacity improvement of UDN. This problem is inherent to the cell-centric network paradigm, and cannot be efficiently solved \cite{Nayebi'15}.
\par
To address the issue, a novel \textit{user-centric} network paradigm called cell-free network has been recently proposed \cite{Nayebi'15}. Unlike the classical cell-centric design principle, the cell-free network utilizes the user-centric transmission design, where all \ac{bs}s in the network jointly serve all users cooperatively without cell boundaries. Due to the efficient cooperation among all distributed \ac{bs}s \cite{Interdonato'19}, the inter-cell interference can be effectively alleviated, and thus the network capacity can be increased accordingly. This promising technique has been considered as a potential candidate for future communication systems \cite{Ngo'17}, and has attracted the increasing research interest such as resource allocation \cite{Mosleh'19}, precoding/beamforming \cite{Attarifar'19}, channel estimation \cite{Jin'19} in recent years.
\par
However, to improve the network capacity further, the deployment of more distributed \ac{bs}s requires high cost and power consumption in the cell-free network. Fortunately, the emerging new technique called \ac{ris} is able to provide an energy-efficient alternative to enhance the network capacity. Equipped with a low-cost, energy-efficient and high-gain metasurface, \ac{ris} is becoming a promising smart radio technique for future 6G communications \cite{Liang'19}. With a large number of low-cost passive elements, \ac{ris} is able to reflect the elertromagentic incident signals to any directions with high array gains by adjusting the phase shifts of its elements \cite{Basar'19}. Since the wireless environment can be effectively manipulated with low cost and energy consumption \cite{Huang'18'2}, \ac{ris} can be used to improve channel capacity \cite{Ntontin'19}, reduce transmit power \cite{Wu'19'2}, enhance transmission reliability \cite{Wang'19}, and enlarge wireless coverage \cite{Pan'19}.
\subsection{Prior works}
The existing research works about \ac{ris} include antenna design \cite{huang2017reconfigurable}, physical model \cite{Ozdogan'19}, channel estimation \cite{Huchen}, joint precoding/beamforming \cite{Nadeem'19}, and etc. Particularly, the prototype of RIS-based wireless communication has been recently developed in \cite{LinglongDai} to demonstrate its functions.
\par
One key guarantee for \ac{ris} to improve the network capacity is the joint precoding. Different from the conventional precoding at the \ac{bs} only, the joint precoding in \ac{ris}-based wireless systems refers to the joint design of the beamforming vector at the \ac{bs} and the phase shifts of the \ac{ris} elements. Different RIS-based scenarios have been studied to maximize the capacity in the literature. Specifically, the authors in \cite{Wu'18} considered a scenario where one \ac{bs} and one \ac{ris} jointly serve a single user, which was an early attempt to realize the capacity enhancement by using low-cost and energy-efficient \ac{ris}. In \cite{Huang'18}, the authors considered a multi-user scenario and maximized the sum-rate of all users. To obtain the cooperation gain, the multi-\ac{bs} scenario was considered in \cite{Pan'19}, while the multi-\ac{ris} case was investigated in \cite{Wang'19}.
\par
Apart from the capacity maximization, the design goal for joint precoding can be different. Specifically, to reduce power consumption, some researchers have proposed the methods to minimize the transmit power \cite{Wu'19'2}, and the authors in \cite{Huang'18'2} have developed a method to maximize the energy efficiency. Some researchers have considered the fairness among users, and some methods have been proposed to maximize the minimum \ac{sinr} \cite{Nadeem'19}. Furthermore, \ac{ris} has been combined with different techniques in the literature. For instance, the RIS-aided \ac{mmwave} system was considered in \cite{Wang'19} and \cite{Di'19}, where the authors discussed the \ac{snr} maximization problem and the joint hybrid precoding design, respectively. In \cite{Li'19} and \cite{Yang'19}, the application of \ac{ris} in wideband \ac{ofdm} systems was discussed to maximize the sum-rate. 
\subsection{Our contributions}
To address the challenge of cell-free network as mentioned above, in this paper we consider to exploit \ac{ris}s to realize the improvement of network capacity with low cost and power consumption\footnote{Simulation codes are provided to reproduce the results presented in this article: http://oa.ee.tsinghua.edu.cn/dailinglong/publications/publications.html.}. Specifically, the contributions of this paper can be summarized as follows.
\begin{itemize}
	\item We propose the concept of RIS-aided cell-free network to further improve the network capacity of the cell-free network with low cost and power consumption. The key idea is to replace some of the required \ac{bs}s in cell-free network by the energy-efficient \ac{ris}s and deploy more \ac{ris}s in the system for capacity enhancement. In the proposed RIS-aided cell-free network, all \ac{bs}s and \ac{ris}s are simultaneously serving all users cooperatively. To the best of our knowledge, this is the first attempt to introduce \ac{ris} in cell-free networks.
	\item For the proposed RIS-aided cell-free network, in a typical wideband scenario, we formulate the problem of joint precoding design at the BSs and RISs to maximize the \ac{wsr} of all users to improve the network capacity. Since the considered scenario is very general, i.e., multiple antennas, multiple \ac{bs}s, multiple \ac{ris}s, multiple users, and multiple carriers, most of the considered scenarios in existing works, such as single \ac{bs}, single \ac{ris}, single user, and single carrier, or some of them are mutiple, are all special cases of the considered scenario in this paper. 
	\item We propose a joint active and passive precoding framework to solve the formulated problem. Specifically, the proposed framework is an alternating optimization algorithm based on the assumption of fully-known channel state information (CSI), which can gradually approximate a feasible solution to joint precoding design. We first decouple the active precoding at the \ac{bs}s and the passive precoding at the \ac{ris}s via Lagrangian dual reformulation and {\it Multidimensional Complex Quadratic Transform (MCQT)}, and the decoupled problem can be reformulated as two \ac{qcqp} subproblems. Then, by solving the two subproblems alternatively, the system \ac{wsr} will finally converge to a feasible solution. 
	\item Challenged by the high-dimensional channels introduced by RISs, acquiring all RIS-aided channels constantly is usually unrealistic. To tackle this issue, by exploiting the property that RISs far from users have little contribution to capacity improvement, the proposed joint precoding framework is further extended to a more practical two-timescale scheme, which can reduce the CSI required by joint precoding from the long-term perspective. Specifically, at the beginning of a large timescale, each user is matched with several well-performed RISs by the proposed {\it linear conic relaxation (LCR)}-based method. Then, in later several small timescales, only the RIS-aided channels of the matched user-RIS pairs are acquired and utilized for joint precoding design, while those of the unmatched pairs are temporarily ignored. Finally, in the next large timescale, the above process will be repeated. 
	\item Simulation results demonstrate that \ac{ris}s can improve the cell-free network capacity significantly. In particular, with limited CSI knowledge, the two-timescale extension of the proposed framework can improve the capacity efficiently with little performance loss. Besides, it is worth noting that, thanks to the generality of the studied problem, the proposed joint precoding framework can also serve as a general solution to maximize the \ac{wsr} in most of the existing RIS-aided scenarios in the literature.	
\end{itemize}
\subsection{Organization and notation}
\textit{Organization:} The rest of the paper is organized as follows. The system model of the proposed RIS-aided cell-free network and corresponding \ac{wsr} maximization problem formulation of joint precoding design are discussed in Section \ref{sec:sys}. The joint precoding framework to solve the formulated problem is proposed in Section \ref{sec:Alg}, and the two-timescale extension of the framework is proposed in Section \ref{sec:Ex}. More framework supplements including convergence and complexity analysis are given in Section \ref{sec:FS}. Simulation results are provided in Section \ref{sec:NSR} to validate the performance of the proposed RIS-aided cell-free network. Finally, in Section \ref{sec:con}, conclusions are drawn and future works are discussed.
\par
\textit{Notations:} $\mathbb{C}$, $\mathbb{R}$, and $\mathbb{R}^{+}$ denote the set of complex, real, and positive real numbers, respectively; ${[\cdot]^{-1}}$, ${[\cdot]^{*}}$, ${[\cdot]^{T}}$, and ${[\cdot]^{H}}$ denote the inverse, conjugate, transpose, and conjugate-transpose operations, respectively; ${[\cdot]^{+}}$ denotes the negative truncation operation, i.e., ${[x]^{+}}=\max\{x,0\}$; ${\left[{\bf v}\right]_{i}}$ is the $i$th element of vector ${\bf v}$; ${\left[{\bf M}\right]_{i,j}}$ is the element at the $i$th row and $j$th column of matrix ${\bf M}$;  $\|\cdot\|$ denotes the Euclidean norm of its argument; $\mathbb{E}\{\cdot\}$ is the expectation operator; ${\rm diag}(\cdot)$ denotes diagonal operation; $\mathfrak{R}\{\cdot\}$ denotes the real part of its argument; ${\rm Tr}\{\cdot\}$ denotes the trace of its argument; $\otimes$ denotes the Kronecker product; $\angle[\cdot]$ denotes the angle of its complex argument; $\ln(\cdot)$ denotes natural logarithm; $\mathbf{I}_{L}$ is an $L\times L$ identity matrix, and $\mathbf{0}_{L}$ is an $L\times L$ zero matrix;  Finally, $\mathbf{e}_{l}$ is an elementary vector with a one at the $l$-th position, and $\mathbf{1}_{L}$ indicates an $L$-length vector with all elements are 1.

\section{System Model of The Proposed RIS-Aided Cell-Free Network}\label{sec:sys}
To improve the network capacity with low cost and power consumption, in this paper we first propose the concept of RIS-aided cell-free network. In this section, the architecture of the proposed RIS-aided cell-free network will be introduced at first. Then, we will discuss the transmitters, channels, and receivers, respectively. Finally, we will formulate the problem of capacity maximization in a wideband RIS-aided cell-free network.
\begin{figure}[!t]
	\centering
	\includegraphics[width=3.4in]{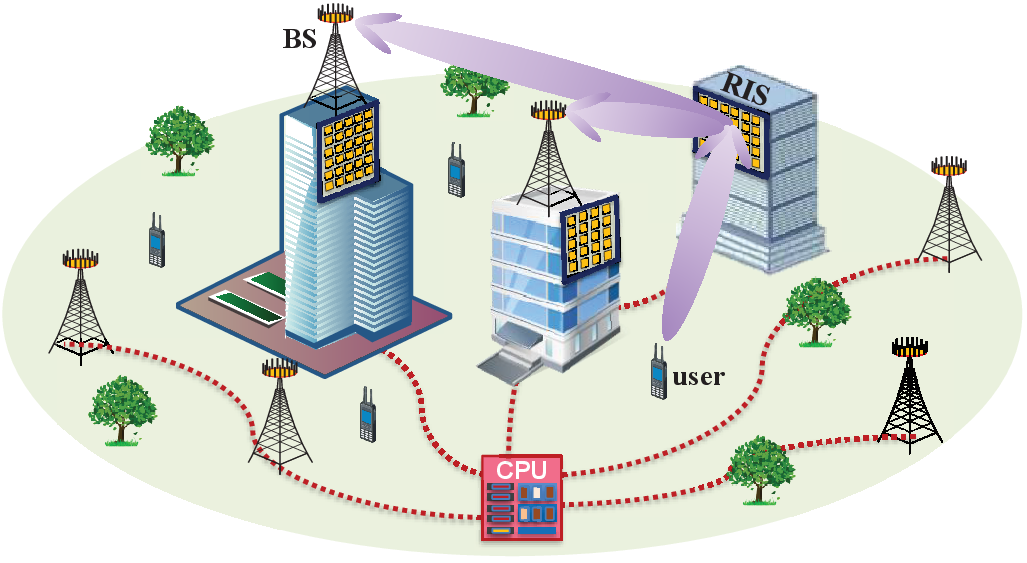}
	\caption{The proposed concept of RIS-aided cell-free network.}
	\label{img:Framework}
	\vspace{-0.8em}
\end{figure}

\subsection{System architecture}
In this paper, we consider a wideband RIS-aided cell-free network as shown in Fig. \ref{img:Framework}, where multiple distributed \ac{bs}s and \ac{ris}s are deployed to cooperatively serve all users. A central processing unit (CPU) is deployed for control and planning, to which all \ac{bs}s are connected by optical cables or wireless backhaul \cite{Siddique'17}. All \ac{ris}s are controlled by the CPU or BSs by wired or wireless control. Particularly, the considered network consists of $B$ \ac{bs}s, $R$ \ac{ris}s, and $K$ multi-antenna users. The number of antennas at the $b$-th \ac{bs} and that at the $k$-th user are $M_b$ and $U_k$, respectively. The number of elements at the $r$-th \ac{ris} is $N_r$. For simplicity but without loss of generality, we assume $M_b$, $U_b$ and $N_r$ are equal to $M$, $U$ and $N$, respectively. Finally, the multi-carrier transmission is considered and the number of available subcarriers is $P$. Let ${\cal N} = \left\{ {1, \cdots ,N} \right\}$, ${\cal B} = \left\{ {1, \cdots ,B} \right\}$, ${\cal R} = \left\{ {1, \cdots ,R} \right\}$, ${\cal K} = \left\{ {1, \cdots ,K} \right\}$ and ${\cal P} = \left\{ {1, \cdots ,P} \right\}$ denote the index sets of \ac{ris} elements, \ac{bs}s, \ac{ris}s, users, and subcarriers, respectively.

\subsection{Transmitters}
In the proposed RIS-aided cell-free network, all \ac{bs}s are synchronized, which is necessary to serve all users by coherent joint transmission \cite{Interdonato'19}. Let ${{\bf{s}}_{p}} \triangleq {\left[ {{s_{p,1}}, \cdots ,{s_{p,K}}} \right]^T} \in {{\mathbb C}^K}$, where ${s}_{p,k}$ denotes the transmitted symbol to the $k$-th user on the $p$-th subcarrier. We assume that the transmitted symbols have normalized power, i.e., $\mathbb{E}\left\{\mathbf{s}_{p}\mathbf{s}_{p}^{H}\right\}=\mathbf{I}_{K}, \forall p \in \mathcal{P}$. In the downlink, the frequency-domain symbol ${s}_{p,k}$ is firstly precoded by the precoding vector ${\bf w}_{b,p,k}\in{{\mathbb C}^{M}}$ at the $b$-th \ac{bs}, so the precoded symbol ${{\bf{x}}_{b,p}}$ at the $b$-th \ac{bs} on the $p$-th subcarrier can be written as
\begin{equation}
	\label{eqn:1}
	{{\bf{x}}_{b,p}} = \sum\nolimits_{k = 1}^K {{{{\bf{w}}_{b,p,k}}{s_{p,k}}}}.
\end{equation}
Then, by \ac{idft}, the frequency-domain signal $\left\{ {{{\bf{x}}_{b,p}}} \right\}_{p = 1}^P$ on all $P$ subcarriers at the $b$-th BS is converted to the time domain. After adding the \ac{cp}, the signal is up-converted to the radio frequency (RF) domain via ${M}$ RF chains of the $b$-th \ac{bs}.
\subsection{Channels}\label{channel}
\begin{figure}[!t]
	\centering
	\includegraphics[width=3.3in]{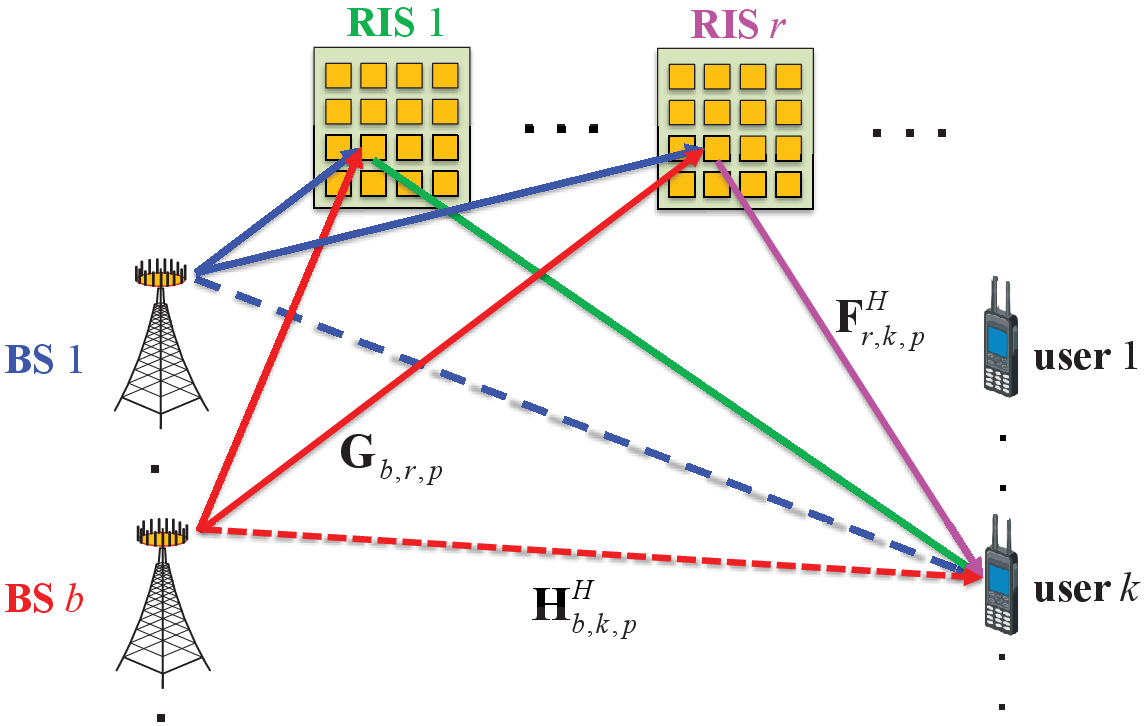}
	\caption{The downlink channels in the wideband RIS-aided cell-free network.}
	\label{img:scenario_show}
	\vspace{-0.8em}
\end{figure}
Thanks to the directional reflection supported by $R$ \ac{ris}s as shown in Fig. \ref{img:scenario_show}, the channel between each \ac{bs} and each user in the proposed RIS-aided cell-free network consists of two parts: the BS-user link and $R$ BS-RIS-user links, where each BS-RIS-user link can be further divided into a BS-RIS link and a RIS-user link. The signal reflection on the \ac{ris}s can be modeled by multiplying the incident signal with a phase shift matrix and forwarding the phase shifted signal to the user \cite{Wu'18}. Therefore, the equivalent channel ${{\bf{h}}_{b,k,p}^H}\in\mathbb{C}^{U\times M}$ from the $b$-th \ac{bs} to the $k$-th user on the $p$-th subcarrier can be written as\footnote{ The signals reflected by the RISs twice and more are ignored, since they are
	much weak due to the large path loss of multiple hops \cite{Han'20}.} \cite{Wu'19}
\begin{equation}
	\label{eqn:2}
	{{\bf{h}}_{b,k,p}^H} = \underbrace {{{\bf{H}}_{b,k,p}^H}}_{\text{BS-user link}} + \underbrace {\sum\limits_{r = 1}^R {{{\bf{F}}_{r,k,p}^H{\bf{\Theta }}_r^H{{\bf{G}}_{b,r,p}}}} }_{\text{BS-RIS-user links}},
\end{equation}
where ${{\bf{H}}_{b,k,p}^H}\in\mathbb{C}^{U\times M}$, ${\bf{G}}_{b,r,p}\in\mathbb{C}^{N\times M}$, and ${\bf{F}}_{r,k,p}^H\in\mathbb{C}^{U\times N}$ denote the frequency-domain channel on the subcarrier $p$ from the \ac{bs} $b$ to the user $k$, from the \ac{bs} $b$ to the \ac{ris} $r$, and from the \ac{ris} $r$ to the user $k$, respectively; ${\bf{\Theta }}_r\in\mathbb{C}^{N\times N}$ denotes the phase shift matrix at the \ac{ris} $r$, which is written as
\begin{equation}
	\label{eqn:3}
	{{\bf{\Theta }}_r} \buildrel \Delta \over = {\mathop{\rm diag}\nolimits} \left( {{\theta _{r,1}}, \cdots ,{\theta _{r,N}}} \right),~\forall r \in \mathcal{R},
\end{equation}
where ${{\theta _{r,n}}}\in{\cal F}$. Note that ${\cal F}$ is the feasible set of the \ac{rc} at \ac{ris}. To study a generalized model, here we assume ${\cal F}$ is the ideal \ac{ris} case, i.e., both the amplitude and the phase of ${\theta _{r,n}}$ associated with the \ac{ris} element can be controlled independently and continuously \cite{Zheng'18}, i.e.,
\begin{equation}
	\label{eqn:4}
	{{\mathcal F}} \triangleq \left\{ {\theta _{r,n}}{\Big |} \left| {{\theta _{r,n}}} \right| \le 1\right\} ,~\forall r \in \mathcal{R},~\forall n\in {{\mathcal N}}.
\end{equation}
Note that the more practical RIS reflection coefficients such as low-resolution discrete phase shifts will be discussed in Subsection \ref{sec:Alg:p5} later.
\subsection{Receivers}
After passing through the equivalent channel ${{\bf{h}}_{b,k,p}^H}$ as s in (\ref{eqn:2}), the signals will be received by the users. The time-domain signals received by the users are down-converted to the baseband at first. After the \ac{cp} removal and the \ac{dft}, the frequency-domain symbols can be finally recovered. Let ${{\bf y}_{b,k,p}}\in\mathbb{C}^U$ denote the baseband frequency-domain signal, which reaches the user $k$ on the subcarrier $p$ from the BS $b$. Then, according to the channel model above, ${{\bf y}_{b,k,p}}$ can be expressed by combining (\ref{eqn:1}) and (\ref{eqn:2}) as
\begin{align}
	\label{eqn:5}
	{{\bf y}_{b,k,p}} & =
	{\bf{h}}_{b,k,p}^H {{\bf{x}}_{b,p}}
	\\&= \left({\bf{H}}_{b,k,p}^H + \sum\limits_{r = 1}^R {{\bf{F}}_{r,k,p}^H{\bf{\Theta }}_r^H{{\bf{G}}_{b,r,p}}} \right)\sum\limits_{j = 1}^K {{{\bf{w}}_{b,p,j}}{s_{p,j}}}. \notag
\end{align}
Since there are $B$ \ac{bs}s serving $K$ users simultaneously, the received signal at user $k$ is the superposition of the signals transmitted by $B$ \ac{bs}s. Let ${{\bf y}_{k,p}}\in\mathbb{C}^U$ denote the received signal at the user $k$ on the subcarrier $p$. Thereby, considering the \ac{awgn} at the receiver, we have the expression of ${{\bf y}_{k,p}}$ as shown in (\ref{eqn:6}) at the bottom of this page,
\begin{figure*}[!b]
	\hrulefill
	\begin{equation}
		\label{eqn:6}
		\begin{aligned}
			{{\bf y}_{k,p}} =& \sum\limits_{b = 1}^B {{{\bf y}_{b,k,p}}}+{\bf z}_{k,p}  =\sum\limits_{b = 1}^B {\sum\limits_{j = 1}^K {\left({\bf{H}}_{b,k,p}^H + \sum\limits_{r = 1}^R {{\bf{F}}_{r,k,p}^H{\bf{\Theta }}_r^H{{\bf{G}}_{b,r,p}}} \right){{\bf{w}}_{b,p,j}}{s_{p,j}}} }+{\bf z}_{k,p}\\=
			& \underbrace {\sum\limits_{b = 1}^B {\left({\bf{H}}_{b,k,p}^H + \sum\limits_{r = 1}^R {{\bf{F}}_{r,k,p}^H{\bf{\Theta }}_r^H{{\bf{G}}_{b,r,p}}} \right){{\bf{w}}_{b,p,k}}{s_{p,k}}} }_{\text{Desired signal to user $k$}}  + \underbrace {\sum\limits_{b = 1}^B {\sum\limits_{j = 1,j \ne k}^K  \left({\bf{H}}_{b,k,p}^H + \sum\limits_{r = 1}^R {{\bf{F}}_{r,k,p}^H{\bf{\Theta }}_r^H{{\bf{G}}_{b,r,p}}} \right){{\bf{w}}_{b,p,j}}{s_{p,j}}}  }_{\text{ Interference from other users}}+ {{\bf z}_{k,p}}
		\end{aligned}.
	\end{equation}
\end{figure*}
where ${\bf z}_{k,p}\triangleq{\left[ {z_{k,p,1}^T, \cdots ,z_{k,p,U}^T} \right]^T}$ denotes the \ac{awgn} with zero mean $\mathbf{0}_{U}$ and covariance ${\bf{\Xi }}_{k,p}=\sigma^2 {{\bf{I}}_U}$. Note that the first term on the right-hand side of (\ref{eqn:6}) is the desired signal to user $k$, while the second term denotes the interference from other users. As there are $P$ subcarriers available in total, we denote the received signal at user $k$ as $\left\{ {{{\bf y}_{k,p}}} \right\}_{p = 1}^P$.
\subsection{Problem fomulation}
Based on the system model above, we consider to maximize the \ac{wsr} of the proposed RIS-aided cell-free network subject to the transmit power constraint at BSs and RC constraint RISs in this subsection. At first, the received signal ${{\bf y}_{b,k,p}}$ in (\ref{eqn:6}) can be simplified as
\begin{align}\label{eqn:7}
	{{\bf{y}}_{k,p}} \notag 
	\mathop { = }\limits^{(a)} &\sum\limits_{b = 1}^B {\sum\limits_{j = 1}^K {\left( {{\bf{H}}_{b,k,p}^H + {\bf{F}}_{k,p}^H{\bf{\Theta }}^H{{\bf{G}}_{b,p}}} \right){{\bf{w}}_{b,p,j}}{s_{p,j}}} }  + {{\bf{z}}_{k,p}} \notag\\
	\mathop { = }\limits^{(b)} &\sum\limits_{b = 1}^B {\sum\limits_{j = 1}^K {{\bf{h}}_{b,k,p}^H{{\bf{w}}_{b,p,j}}{s_{p,j}}} }  + {{\bf{z}}_{k,p}} \\
	\mathop { = }\limits^{(c)} &\sum\limits_{j = 1}^K {{\bf{h}}_{k,p}^H{{\bf{w}}_{p,j}}{s_{p,j}}}  + {{\bf{z}}_{k,p}}, \notag
\end{align}
where $(a)$ holds by defining ${\bf{\Theta }} = {\rm diag}\left({{\bf{\Theta }}_1}, \cdots ,{{\bf{\Theta }}_R}\right)$, ${{\bf{F}}_{k,p}} = {\left[ {{\bf{F}}_{1,k,p}^T, \cdots ,{\bf{F}}_{R,k,p}^T} \right]^T}$, and ${{\bf{G}}_{b,p}} = {\left[ {{\bf{G}}_{b,1,p}^T, \cdots ,{\bf{G}}_{b,R,p}^T} \right]^T}$, $(b)$ holds according to (\ref{eqn:2}), and $(c)$ holds by defining ${\bf{h}}_{k,p}=\left[{\bf{h}}_{1,k,p}^T,\cdots,{\bf{h}}_{B,k,p}^T\right]^T$ and ${\bf{w}}_{p,k}=\left[{\bf{w}}_{1,p,k}^T,\cdots,{\bf{w}}_{B,p,k}^T\right]^T$. Then, the \ac{sinr} for the transmitted symbol $s_{p,k}$ at the user $k$ on the subcarrier $p$ can be easily calculated as
\begin{align}\label{eqn:8}
	&{\gamma _{k,p}} \\=& {\bf{w}}_{p,k}^H{{\bf{h}}_{k,p}}\!{\left( {\sum\limits_{j = 1,j \ne k}^K \!\!\! {{\bf{h}}_{k,p}^H{{\bf{w}}_{p,j}}{{\left( {{\bf{h}}_{k,p}^H{{\bf{w}}_{p,j}}} \right)}^H}} \!\!+\! {{\bf{\Xi }}_{k,p}}} \right)^{ - 1}}\!\!{\bf{h}}_{k,p}^H{{\bf{w}}_{p,k}}. \notag
\end{align}
Thereby, the \ac{wsr} $R_{\rm sum}$ of all $K$ users is given by
\begin{equation}
	\label{eqn:9}
	\begin{aligned}
		R_{\rm sum} =
		\sum\limits_{k = 1}^K {\sum\limits_{p = 1}^P {\eta _k}{{{\log }_2}\left( {1 + {\gamma_{k,p}}} \right)} },
	\end{aligned}
\end{equation}
where $\eta_{k}\in\mathbb{R}^{+}$ represents the weight of the user $k$ and $R_{k,p}$ denotes the rate of user $k$ on subcarrier $p$.
\par
Finally, the \ac{wsr} maximization optimization problem  can be originally formulated as
\begin{subequations}\label{eqn:11}
	\begin{align}
		\label{eqn:11a}
		{\cal P}^{\rm o}:~&\mathop {{\rm{max}}}\limits_{{\bf{\Theta }},{\bf{W}}} ~{	R_{\rm sum}}({\bf{\Theta }},{\bf{W}}){\rm{ = }}\sum\limits_{k = 1}^K {\sum\limits_{p = 1}^P {{\eta_k}{{\log }_2}\left( {1 + {\gamma_{k,p}}} \right)} }\\
		\label{eqn:11b}
		&\,{\rm{s.t.}}~~~C_1:\sum\limits_{k = 1}^K {\sum\limits_{p = 1}^P {\left\| {{{\bf{w}}_{b,p,k}}} \right\|}^2 }  \le {P_{b,\max }},~\forall b \in \mathcal{B},\!\!\\
		\label{eqn:11c}
		&~~~~~~~\,C_2:{{\theta _{r,n}}}\in{\cal F},~\forall r\in{{\cal R}},\forall n\in{{\cal N}},
	\end{align}
\end{subequations}
where ${P_{b,\max }}$ denotes the maximum transmit power of the \ac{bs} $b$ and we have defined ${\bf{W}}$ as follows for expression simplicity:
\begin{equation}
	\begin{aligned}
		{\bf{W}} \!=\! {\left[{\bf{w}}_{1,1}^T,{\bf{w}}_{1,2}^T, \cdots ,{\bf{w}}_{1,K}^T,{\bf{w}}_{2,1}^T,{\bf{w}}_{2,2}^T, \cdots ,{\bf{w}}_{P,K}^T\right]^T}.
	\end{aligned}
\end{equation}
\par
Due to the non-convex complex objective function (\ref{eqn:11a}), the joint optimization of the phase shift matrix  $\bf{\Theta }$ and the precoding vector ${\bf{W}}$ is very challenging. Fortunately, inspired by the fractional programming (FP) methods, we propose a  joint precoding framework to find a feasible solution to the problem ${\cal P}^{\rm o}$ in the following Section \ref{sec:Alg}. 

\section{Proposed Joint Precoding Framework}\label{sec:Alg}
In this section, we present the proposed joint precoding framework to solve the \ac{wsr} optimization problem ${\cal P}^{\rm o}$ in (\ref{eqn:11}). Specifically, the section is summarized as follows. An overview of the proposed framework is first provided in Subsection \ref{sec:Alg:p1}, where the problem ${\cal P}^{\rm o}$ in (\ref{eqn:11}) is divided into three subproblems. Then, the detailed algorithms to solve these three subproblems are given in Subsections \ref{sec:Alg:p2}, \ref{sec:Alg:p3}, and \ref{sec:Alg:p4}, respectively. 
\subsection{Overview of the proposed joint precoding framework}\label{sec:Alg:p1}
\begin{figure}[!t]
	\centering
	\includegraphics[width=3.4in]{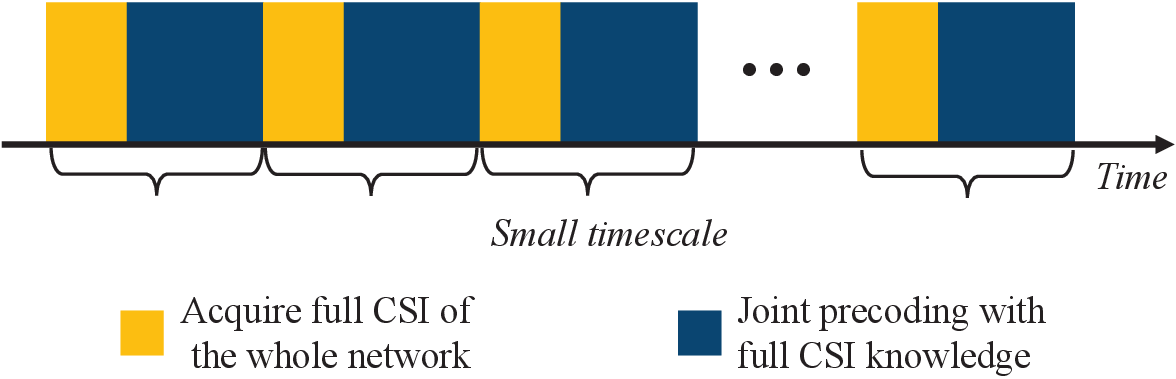}
	\caption{The dynamical working process of the proposed joint precoding framework over time.}
	\label{img:structure1}
	\vspace{-0.8em}
\end{figure}
As the basis of the joint precoding design, we assume that the CSI of the whole RIS-aided cell-free network can be fully acquired by the CPU in advance. Then, based on the fully-known CSI, the joint precoding at the BSs and RISs are further designed and employed. Intuitively, we draw Fig. \ref{img:structure1} to show this dynamical process over time. Under the assumption of fully-known CSI in each small timescale, we focus on solving the problem ${\cal P}^{\rm o}$ in (\ref{eqn:11}) to find a feasible precoding design ${\bf W}^{\rm opt}$ and ${\bm \Theta}^{\rm opt}$ for each small timescale as follows.
\par
At first, to deal with the complexity of sum-logarithms in the \ac{wsr} maximization problem ${\cal P}^{\rm o}$ in (\ref{eqn:11}), by utilizing Lagrangian dual reformulation (LDR), a method has been proposed in \cite{Shen'18'1} to decouple the logarithms. Based on this, we have the following \textit{Proposition 1}.
\begin{proposition}\label{proposition:1}
	By introducing an auxiliary variable ${\bm \rho }\in \mathbb{R}^{PK}$ with ${\bm \rho } = {[{\rho _{1,1}},\rho _{1,2}^{}, \cdots ,\rho _{1,K},\rho _{2,1}^{},\rho _{2,2}^{}, \cdots ,\rho _{P,K}^{}]^T}$, the original problem ${{\cal P}^{\rm o}}$ in (\ref{eqn:11}) is equivalent to
	\begin{align}
		{\bar {\cal P}}:~~&\mathop {{\rm{max}}}\limits_{{\bf{\Theta }},{\bf{W}},{\bm \rho}}~~~f({\bf{\Theta }},{\bf{W}},{\bm{\rho }}) \notag \\
		&\,{\rm{s.t.}}~~~C_1:\sum\limits_{k = 1}^K {\sum\limits_{p = 1}^P {\left\| {{{\bf{w}}_{b,p,k}}} \right\|}^2 }  \le {P_{b,\max }},~\forall b \in \mathcal{B},\! \label{eqn:13}\\
		&~~~~~~~\,C_2:{{\theta _{r,n}}}\in{\cal F},~\forall r\in{{\cal R}},\forall n\in{{\cal N}}, \notag
	\end{align}
	where the new objective function $f({\bf{\Theta }},{\bf{W}},{\bm{\rho }})$ is
	\begin{align}\label{eqn:14}
		f({\bf{\Theta }},{\bf{W}},{\bm{\rho }}) =&\sum\limits_{k = 1}^K {\sum\limits_{p = 1}^P {{\eta _k}{{\ln }}\left( {1 + {\rho _{k,p}}} \right)}} - \sum\limits_{k = 1}^K {\sum\limits_{p = 1}^P {{\eta _k}{\rho _{k,p}}} } \notag  \\&+  \sum\limits_{k = 1}^K {\sum\limits_{p = 1}^P {{\eta _k}(1 + {\rho _{k,p}}){f_{k,p}}({\bf{\Theta }},{\bf{W}})} } ,
	\end{align}
	wherein the function $f_{k,p}({\bf{\Theta }},{\bf{W}})$ is denoted by
	\begin{align}\label{eqn:15}	
		&{f_{k,p}}({\bf{\Theta }},{\bf{W}}) =\\& {\bf{w}}_{p,k}^H{{\bf{h}}_{k,p}}{\left( {\sum\limits_{j = 1}^K {{\bf{h}}_{k,p}^H{{\bf{w}}_{p,j}}{{\left( {{\bf{h}}_{k,p}^H{{\bf{w}}_{p,j}}} \right)}^H}}  + {{\bf{\Xi }}_{k,p}}} \right)^{ - 1}}{\bf{h}}_{k,p}^H{{\bf{w}}_{p,k}}. \notag
	\end{align}
\end{proposition}
\par
\begin{algorithm}[!t] 
	\caption{Proposed Joint Precoding Framework.} 
	\label{alg:1} 
	\begin{algorithmic}[1] 
		\REQUIRE ~~ 
		All channels ${{\bf{H}}_{b,k,p}},{\bf{G}}_{b,r,p}$ and $ {\bf{F}}_{r,k,p}$ where $\forall b\in {\cal B},k\in {\cal K},p\in {\cal P}$.
		\ENSURE ~~ 
		Optimized active precoding vector $\bm{W}$; Optimized passive precoding matrix $\bm{\Theta}$; Weighted sum-rate $R_{{\rm sum}}$.	
		\STATE Initialize $\bm{W}$ and $\bm{\Theta}$;
		\WHILE {no convergence of $R_{{\rm sum}}$}
		\STATE Update ${\bm \rho}$ by (\ref{eqn:16});
		\STATE Update ${\bm \xi}$ by (\ref{eqn:20});
		\STATE Update ${\bf W}$ by solving (\ref{eqn:25});
		\STATE Update ${\bm \varpi}$ by (\ref{eqn:31});
		\STATE Update ${\bm{\Theta}}$ by solving (\ref{eqn:37});
		\ENDWHILE	
		\RETURN $\bf{W}^{\rm opt}$, ${\bm{\Theta}}^{\rm opt}$, and $R_{{\rm sum}}$. 
	\end{algorithmic}
\end{algorithm}
Then, we propose the joint active and passive precoding framework to optimize the variables ${\bm{\rho }}$, ${\bf{W}}$, and ${\bf{\Theta }}$ in (\ref{eqn:13}) iteratively. After introducing two auxiliary variables ${\bm \xi }$ and $\bm \varpi$, the proposed joint precoding framework to maximize the \ac{wsr} is summarized in {\bf Algorithm 1}. In this framework, the variables ${\bm \rho}$, ${\bm \xi }$, ${\bf W}$, ${\bm \varpi }$, and ${\bm {\Theta} }$ are alternately updated until the convergence of the objective function is achieved.

The optimal solutions to these variables at each step will be introduced in the following three subsections. Specifically, the solution to ${\bm \rho}^{\rm opt}$ is firstly present in Subsection \ref{sec:Alg:p2}. Then, the solutions to ${\bm \xi }^{\rm opt}$ and ${\bf W}^{\rm opt}$ for the active precoding design are provided in Subsection \ref{sec:Alg:p3}. After that, the solutions to ${\bm \varpi }^{\rm opt}$ and ${\bm {\Theta} }^{\rm opt}$ for the passive precoding design are finally discussed in Subsection \ref{sec:Alg:p4}. 
\subsection{Fix $({\bf{\Theta}},{\bf{W}})$ and solve ${\bm \rho}^{\rm opt}$}\label{sec:Alg:p2}
Given fixed $({\bf{\Theta }^\star},{\bf{W}^\star})$, the optimal $\bm \rho$ in (\ref{eqn:14}) can be obtained by solving $\partial f / \partial \rho _{k,p}=0$ for $\forall k \in \mathcal{K},\forall p \in \mathcal{P}$. The solution can be written as
\begin{equation}
	\label{eqn:16}
	\begin{aligned}
		\rho _{k,p}^{\rm opt} = {\gamma_{k,p}^\star},~~\forall k \in \mathcal{K},\forall p \in \mathcal{P}.
	\end{aligned}
\end{equation}
By substituting $\rho _{k,p}^{\rm opt}$ in (\ref{eqn:16}) back into $f$ in (\ref{eqn:14}), one can notice that, only the last term in (\ref{eqn:14}) is associated with the variables ${\bf{W}}$ and ${\bf{\Theta }}$. Hence, the problem ${\bar {\cal P}}$ in (\ref{eqn:13}) can be further solved as shown in the following two subsections.
\subsection{Active precoding: fix $({\bf{\Theta }},{\bm{\rho}})$ and solve ${\bf W}^{\rm opt}$} \label{sec:Alg:p3}
In the case of given $({\bf{\Theta}^\star},{\bm{\rho}^\star})$, the equivalent \ac{wsr} maximization problem $\bar{\cal P}$ in (\ref{eqn:13}) can be reformulated as the following subproblem ${{\cal P}_{ \rm active}}$ for the active precoding design at \ac{bs}s:
\begin{equation}\label{eqn:17}
	\begin{aligned}
		{{\cal P}_{ \rm active}:}~&\mathop {{\rm{max}}}\limits_{\bf{W}}~g_{1}({\bf{W}})=\sum\limits_{k = 1}^K {\sum\limits_{p = 1}^P {\mu_{k,p}{f_{k,p}}({\bf{\Theta }^\star},{\bf{W}})} } \\
		&\,{\rm{s.t.}}~C_1:\sum\limits_{k = 1}^K {\sum\limits_{p = 1}^P {\left\| {{{\bf{w}}_{b,p,k}}} \right\|}^2 }  \le {P_{b,\max }},~\forall b \in \mathcal{B},\!\!
	\end{aligned}
\end{equation}
where $\mu_{k,p}={\eta _k}(1 + {\rho _{k,p}^\star})$ holds.  However, note that the reformulated subproblem ${\cal P}_{ \rm active}$ in (\ref{eqn:17}) is still too difficult to solve due to the high-dimensional non-convex  $f_{k,p}$ in (\ref{eqn:15}). Specifically, due to the channels introduced by the RIS-aided cell-free network, this subproblem is actually a special high-dimensional sum-of-fractions problem. Different from the familiar scalar-form fractions, the high-dimensional “fractions” in (\ref{eqn:15}) are the products of matrices and inverse matrices. Thereby, the non-convexity of $f_{k,p}$ in (\ref{eqn:15}) cannot be simply relaxed by adopting the common FP methods such as the Dinkelbach's algorithms \cite{Dinklebach}.
\par
To tackle this issue, we fortunately notice a recently proposed method called {\it multidimensional complex quadratic transform (MCQT)} \cite{Shen'18'1}. Different from the common FP methods, MCQT extends the common scalar-form fractional programming to matrix-form and can be utilized to address the non-convexity of the high-dimensional “fractions” \cite{Shen'18'1}. Since  $f_{k,p}$ in (\ref{eqn:17}) just meets the {\it concave-convex} conditions required by MCQT, we can apply MCQT to reformulate the subproblem (\ref{eqn:17}) to address its non-convexity. In this way, we obtain \textit{Proposition 2} as below.
\begin{proposition}\label{proposition:2}
	Exploiting the fractional programming method MCQT and by introducing auxiliary variables ${\bm \xi }_{p,k}\in \mathbb{C}^{U}$ with ${\bm \xi }= {[{\bm \xi _{1,1}},\bm\xi _{1,2}, \cdots ,\bm\xi _{1,K},\bm\xi _{2,1},\bm\xi _{2,2}, \cdots ,\bm\xi _{P,K}]}$, the subproblem ${{\cal P}_{ \rm active}}$ in (\ref{eqn:17}) can be further reformulated as
	\begin{equation}
		\label{eqn:18}
		\begin{aligned}
			{\bar{\cal P}_{ \rm active}:}~&\mathop {{\rm{max}}}\limits_{\bf{W},{\bm \xi}}~~~g_{2}({\bf{W},{\bm \xi}}) \\
			&\,{\rm{s.t.}}~C_1:\sum\limits_{k = 1}^K {\sum\limits_{p = 1}^P {\left\| {{{\bf{w}}_{b,p,k}}} \right\|}^2 }  \le {P_{b,\max }},~\forall b \in \mathcal{B},\!\!
		\end{aligned}
	\end{equation} 
	where
	\begin{align}\label{eqn:19}
		&{{g}_2}({\bf{W}},{\bm{\xi }}) = \sum\limits_{k = 1}^K {\sum\limits_{p = 1}^P {2\sqrt {{\mu}_{k,p}}\,\,{\mathfrak{R}} \left\{ {\bm{\xi }}_{k,p}^H{\bf{h}}_{k,p}^H{{\bf{w}}_{p,k}}\right\} } } 
		\\&- \sum\limits_{k = 1}^K {\sum\limits_{p = 1}^P {{\bm{\xi }}_{k,p}^H\left(\sum\limits_{j = 1}^K {{\bf{h}}_{k,p}^H{{\bf{w}}_{p,j}}{{\left( {{\bf{h}}_{k,p}^H{{\bf{w}}_{p,j}}} \right)}^H}}  + {{\bf{\Xi }}_{k,p}} \right){{\bm{\xi }}_{k,p}}} } . \notag
	\end{align}
\end{proposition}
\par
Thereby, the updating of ${\bf{W}}$ can be divided into two steps of updating ${\bm{\xi }}$ and  ${\bf{W}}$ in turn. To achieve this, the reformulated subproblem ${\bar{\cal P}_{ \rm active}}$ in (\ref{eqn:18}) can be divided into two subproblems and solved respectively as follows.
\subsubsection{Fix $\bf W$ and solve ${\bm \xi }^{\rm opt}$}
While fixing $\bf W$ in $\bar{\cal P}_{ \rm active}$ in (\ref{eqn:18}), by setting $\partial g_2 / \partial {\bm\xi} _{k,p}$ to zero, the optimal ${\bm \xi }$ can be obtained by
\begin{align}\label{eqn:20}
	{\bm \xi} _{k,p}^{\rm opt} =\! \sqrt {{\mu}_{k,p}} &{\left( \sum\limits_{j = 1}^K {{\bf{h}}_{k,p}^H{{\bf{w}}_{p,j}}{{\left( {{\bf{h}}_{k,p}^H{{\bf{w}}_{p,j}}} \right)}^H}}\!\!+ {{\bf{\Xi }}_{k,p}} \right)^{\!\!- 1}}\!\!{\bf{h}}_{k,p}^H{{\bf{w}}_{p,k}}, \notag\\&
	\forall k \in \mathcal{K}~,\forall p \in \mathcal{P}.
\end{align}
\subsubsection{Fix ${\bm \xi }$ and solve ${\bf W}^{\rm opt}$}
While fixing ${\bm \xi }$ in $\bar{\cal P}_{ \rm active}$ in (\ref{eqn:18}), for simplification and clarity of (\ref{eqn:18}), we can first define
\begin{subequations}\label{eqn:21}
	\begin{align}
		&{{\bf{a}}_p} = \sum\nolimits_{k = 1}^K {{\bf{h}}_{k,p}{{\bm{\xi }}_{k,p}}{\bm{\xi }}_{k,p}^H{{\bf{h}}_{k,p}^H}},\\
		&{{\bf{A}}_p} = {\bf I}_K \otimes {{\bf{a}}_p}, {\quad \quad }{\bf v}_{k,p}= {\bf{h}}_{k,p}{\bm \xi}_{k,p}.
	\end{align}
\end{subequations}
Then, by substituting (\ref{eqn:21}) into $g_2$ in (\ref{eqn:19}), we can rewritten ${{g}_2}$ as
\begin{equation}
	\label{eqn:22}
	{{g}_2}\left({\bf{W}}\right) =   - {{\bf{W}}^H}{\bf{AW}} +{\mathfrak{R}} \left\{2 {{\bf{V}}^H}{\bf{W}}\right\}- Y,
\end{equation} 
where 	
\begin{subequations}
	\begin{align}
		&{{\bf{A}} \!=\! {\rm diag}\left({{\bf{A}}_1},\cdots,{{\bf{A}}_P}\right)},\quad Y \!=\! \sum\limits_{k = 1}^K {\sum\limits_{p = 1}^P {{\bm{\xi }}_{k,p}^H{{\bf{\Xi }}_{k,p}}{{\bm{\xi }}_{k,p}}} },\\
		&{\bf{V}} = {[{{\bf v}^{T}_{1,1}},{\bf v}_{1,2}^{T}, \cdots ,{\bf v}_{1,K}^{T},{\bf v}_{2,1}^{T},{\bf v}_{2,2}^{T}, \cdots ,{\bf v}_{P,K}^{T}]^T}.
	\end{align}
\end{subequations}
Therefore, the active precoding problem ${\bar{\cal P}_{ \rm active}}$ in (\ref{eqn:18}) can be further simplified as
\begin{figure*}[!b]
	\hrulefill
	\begin{equation}
		\label{eqn:27}
		\begin{aligned}
			{g_4}({\bf{\Theta }}) = \sum\limits_{k = 1}^K {\sum\limits_{p = 1}^P {\sqrt {{\mu _{k,p}}} {\bf{Q}}_{k,p,k}^H\left( {\bf{\Theta }} \right){{\left( {\sum\limits_{j = 1}^K {{{\bf{Q}}_{k,p,j}}\left( {\bf{\Theta }} \right){\bf{Q}}_{k,p,k}^H\left( {\bf{\Theta }} \right)}  + {{\bf{\Xi }}_{k,p}}} \right)}^{ - 1}}{{\bf{Q}}_{k,p,k}}\left( {\bf{\Theta }} \right)} } .
		\end{aligned}
	\end{equation}
\end{figure*}
\begin{equation}
	\label{eqn:25}
	\begin{aligned}
		{\hat{\cal P}_{ \rm active}:}~~&\mathop {{\rm{min}}}\limits_{\bf{W}}~~g_{3}({\bf{W}})={{\bf{W}}^H}{\bf{AW}} -{\mathfrak{R}} \left\{2 {{\bf{V}}^H}{\bf{W}}\right\} \\
		&\,{\rm{s.t.}}~~~~C_1:{\bf{W}}^H{{\bf D}_b}{\bf{W}}  \le {P_{b,\max }},~\forall b \in \mathcal{B},
	\end{aligned}
\end{equation} 
where ${{\bf D}_b}={{\bf{I}}_{PK}} \otimes \left\{ \left({{\bf{e}}_b}{\bf{e}}_b^H\right) \otimes {{\bf{I}}_M}\right\}$ with ${\bf{e}}_b\in\mathbb{R}^{B}$. Since the matrices $\bf A$ and ${\bf D}_b$ ($\forall b \in \mathcal{B}$) are all positive semidefinite, the simplified subproblem ${\hat{\cal P}_{ \rm active}}$ in (\ref{eqn:25}) is a standard \ac{qcqp} problem, which can be optimally solved by many existing methods such as alternating direction method of multipliers (ADMM) \cite{admm}. 
\par
However, note that the adoption of ADMM in QCQP problem requires the inversion for the matrix $\bf A$ (along with Lagrange multipliers). Due to the high-dimensional channels of cell-free network, the dimension of $\bf A$  is usually very high ($BMPK$). As a result, the inversion for $\bf A$ has a high computational complexity of about ${\cal O}\left(B^3M^3P^3K^3\right)$, which may prevent the precoding design from practical application. To avoid the matrix inversion operation thus reducing the complexity, here we provide an inversion-free feasible solution by exploiting the {\it primal-dual subgradient (PDS)} method \cite{Boyd'14} to obtain ${\bf W}^{\rm opt}$ in Appendix \ref{appendix:1}.
\subsection{Passive precoding: fix $({\bm{\rho}},{\bf{W}})$ and solve ${\bf \Theta}^{\rm opt}$}\label{sec:Alg:p4}

Based on the given $({\bm{\rho}}^\star,{\bf{W}}^\star)$, for the equivalent \ac{wsr} maximization problem ${\bar {\cal P}}$ in (\ref{eqn:13}), the subproblem of the RIS precoding design at \ac{ris}s can be equivalently rewritten as
\begin{equation}
	\label{eqn:26}
	\begin{aligned}
		{{\cal P}_{ \rm passive}:}~~~&\mathop {{\rm{max}}}\limits_{\bf{\Theta}}~~g_{4}({\bf{\Theta}})=\sum\limits_{k = 1}^K {\sum\limits_{p = 1}^P {\mu_{k,p}{f_{k,p}}({\bf{\Theta }},{\bf{W}^\star})} } \\
		&~{\rm{s.t.}}~~~C_2:{{\theta _{r,n}}}\in{\cal F},~\forall r\in{{\cal R}},\forall n\in{{\cal N}},
	\end{aligned}
\end{equation}
where $\mu_{k,p}={\eta _k}(1 + {\rho _{k,p}^\star})$.
Similarly, to reduce the complexity, we wish to simplify the expression of $g_4$ in (\ref{eqn:26}). Firstly, by defining a new auxiliary function with respect to ${\bf{\Theta}}$ as
\begin{equation}
	\label{eqn:28}
	{{\bf{Q}}_{k,p,j}}\left( {\bf{\Theta }} \right) = \sum\limits_{b = 1}^B {\left( {{\bf{H}}_{b,k,p}^H + {\bf{F}}_{k,p}^H{{\bf{\Theta }}^H}{{\bf{G}}_{b,p}}} \right){{\bf{w}}_{b,p,j}}},
\end{equation}
we can rewrite $g_4$ in (\ref{eqn:26}) as (\ref{eqn:27}) at the bottom of next page.
However, this subproblem is still hard to solve due to the mutidimensional fractions in $f_{k,p}$ in (\ref{eqn:15}). Notice that the subproblem (\ref{eqn:26}) satisfies the {\it concave-convex} conditions \cite{Shen'18'1}, again we exploit the MCQT to address this issue again by using the following \textit{Proposition 3}.
\begin{proposition}\label{proposition:3}
	With the fractional programming method MCQT, by introducing an auxiliary variable ${\bm \varpi }_{p,k}\in \mathbb{C}^{U}$ and ${\bm \varpi }= {[{\bm \varpi _{1,1}},\bm\varpi _{1,2}, \cdots ,\bm\varpi _{1,K},\bm\varpi _{2,1},\bm\varpi _{2,2}, \cdots ,\bm\varpi _{P,K}]}$, the passive precoding subproblem ${{\cal P}_{ \rm passive}}$ in (\ref{eqn:26}) can be reformulated as
	\begin{subequations}\label{eqn:29}
		\begin{align}
			\label{eqn:29a}
			{\bar{\cal P}_{ \rm passive}:}~~
			&\mathop {{\rm{max}}}\limits_{\bf{\Theta}}~~{{g}_5}\left({\bf{\Theta }},{\bm{\varpi }}\right)=\sum\limits_{k = 1}^K {\sum\limits_{p = 1}^P {{g_{k,p}}\left( {\bf{\Theta }},{\bm{\varpi }} \right)} } \\
			&~{\rm{s.t.}}~~C_2:{{\theta _{r,n}}}\in{\cal F},~\forall r\in{{\cal R}},\forall n\in{{\cal N}},
		\end{align}
	\end{subequations}
	where	
	\begin{equation}
		\label{eqn:30}
		\begin{aligned}
			&{{g}_{k,p}}({\bf{\Theta }},{\bm{\varpi }}) = 2\sqrt {{\mu_{k,p}}}\, {\mathfrak{R}} \left\{ {{\bm{\varpi }}_{k,p}^H{{\bf{Q}}_{k,p,k}}\left( {\bf{\Theta }} \right)} \right\} \\&- {\bm{\varpi }}_{k,p}^H\left( \sum\limits_{j = 1}^K {{{\bf{Q}}_{k,p,j}}\left( {\bf{\Theta }} \right){\bf{Q}}_{k,p,j}^H\left( {\bf{\Theta }} \right)}  + {{\bf{\Xi }}_{k,p}} \right){{\bm{\varpi }}_{k,p}}.
		\end{aligned}
	\end{equation} 
\end{proposition}
\par
Next, similar to the previous processing of solving the subproblem ${\bar{\cal P}_{ \rm active}}$ in (\ref{eqn:18}), we consider to optimize two variables ${\bm{\varpi }}$ and ${\bf{\Theta }}$ in (\ref{eqn:29}) in turn. The reformulated subproblem ${\bar{\cal P}_{ \rm passive}}$ in (\ref{eqn:29}) can be further divided into two subproblems and respectively solved as follows.
\subsubsection{Fix $\bm \Theta$ and solve ${\bm \varpi }^{\rm opt}$}
For given fixed ${\bm \Theta}$ in $\bar{\cal P}_{ \rm passive}$ in (\ref{eqn:29}), by solving $\partial g_5 / \partial {\bm\varpi} _{k,p}=0$ for $\forall k \in \mathcal{K}$ and $\forall p \in \mathcal{P}$, we can obtain the optimal ${\bm \varpi }_{k,p}$ for all $k\in{\cal K}$ and $p\in{\cal P}$ by
\begin{align}\label{eqn:31}
	&{\bm{\varpi }}_{k,p}^{\rm opt} =\\& \sqrt {{\mu_{k,p}}} {\left(\sum\limits_{j = 1}^K {{{\bf{Q}}_{k,p,j}}\left( {{{\bf{\Theta }}}} \right){\bf{Q}}_{k,p,j}^H\left( {{{\bf{\Theta }}}} \right)}  + {{\bf{\Xi }}_{k,p}} \right)^{ \!\!\!-1}}\!\!{{\bf{Q}}_{k,p,k}}\left( {\bf{\Theta }} \right).\notag
\end{align}
\subsubsection{Fix $\bm \varpi$ and solve ${\bf \Theta}^{\rm opt}$}
While fixing $\bm \varpi$ in $g_5$ in (\ref{eqn:29}), due to the complexity of $\bar{\cal P}_{ \rm passive}$ in (\ref{eqn:29}), we first consider to simplify the expression of $g_5$ by using the new auxiliary function ${{\bf{Q}}_{k,p,j}}\left( {\bf{\Theta }} \right)$ with respect to ${\bf{\Theta }}$ in (\ref{eqn:28}) as follows:
\begin{align}
	&{\bm\varpi} _{k,p}^H{{\bf{Q}}_{k,p,j}}\left( {\bf{\Theta }} \right) \notag \\&\mathop { = }\limits^{(a)} \sum\limits_{b = 1}^B {\left( {{\bm\varpi} _{k,p}^H{\bf{H}}_{b,k,p}^H{{\bf{w}}_{b,p,j}} + {\bm\varpi} _{k,p}^H{\bf{F}}_{k,p}^H{{\bf{\Theta }}^H}{{\bf{G}}_{b,p}}{{\bf{w}}_{b,p,j}}} \right)} \notag \\
	&\mathop { = }\limits^{(b)} \sum\limits_{b = 1}^B {{\bm\varpi} _{k,p}^H{\bf{H}}_{b,k,p}^H{{\bf{w}}_{b,p,j}}}  \!+\! {{\bm{\theta }}^H}\!\sum\limits_{b = 1}^B {{\rm diag}\left( {{\bm\varpi} _{k,p}^H{\bf{F}}_{k,p}^H} \right){{\bf{G}}_{b,p}}{{\bf{w}}_{b,p,j}}} \notag \\
	\label{eqn:32}
	&\mathop { = }\limits^{(c)} {c_{k,p,j}} + {{\bm{\theta }}^H}{{\bf{g}}_{k,p,j}},
\end{align}
where $(a)$ holds according to (\ref{eqn:27}), $(b)$ is obtained by defining ${\bm{\theta }}={\bm{\Theta }}{\bf 1}_{RN}$, and $(c)$ is achieved by defining 
\begin{subequations}
	\begin{align}
		&{c_{k,p,j}} = \sum\nolimits_{b = 1}^B {{\bm\varpi} _{k,p}^H{\bf{H}}_{b,k,p}^H{{\bf{w}}_{b,p,j}}}, \label{eqn:c_kpj}\\
		&{{\bf{g}}_{k,p,j}} = \sum\nolimits_{b = 1}^B {{\rm diag}\left( {{\bm\varpi} _{k,p}^H{\bf{F}}_{k,p}^H} \right){{\bf{G}}_{b,p}}{{\bf{w}}_{b,p,j}}}.
	\end{align}
\end{subequations}
By substituting (\ref{eqn:32}) into (\ref{eqn:30}), we obtain:
\begin{align}
	\label{eqn:34}	
	&{{{g}}_{k,p}}({\bf{\Theta }}) =2\sqrt {{{\mu}_{k,p}}} \,{\mathfrak{R}} \left\{ {{c_{k,p,k}} + {{\bm{\theta }}^H}{{\bf{g}}_{k,p,k}}} \right\} \\&-\! \sum\limits_{j = 1}^K {\left({c_{k,p,j}} \!+\! {{\bm{\theta }}^H}{{\bf{g}}_{k,p,j}}\right)\left(c_{k,p,j}^* \!+\! {{\bf{g}}^H_{k,p,j}}{\bm{\theta }}\right)}\!-\!{\bm\varpi} _{k,p}^H{{\bf{\Xi }}_{k,p}}{{\bm\varpi} _{k,p}} . \notag
\end{align}
Then, we can further substitute (\ref{eqn:34}) into (\ref{eqn:29a}), so that $g_5$ in (\ref{eqn:29a}) can be simplified as
\begin{equation}
	\label{eqn:35}
	\begin{aligned}
		{{g}_5}({\bm{\Theta}}) =   - {{\bm{\theta}}^H}{\bf{	\Lambda}}{\bm \theta} +{\mathfrak{R}} \left\{2 {{\bm{\theta}}^H}{\bm{\nu}}\right\}- 	\zeta,
	\end{aligned}
\end{equation}
where
\begin{subequations}
	\begin{align}
		{\bm\Lambda } =& \sum\limits_{k = 1}^K {\sum\limits_{p = 1}^P {\sum\limits_{j = 1}^K {{{\bf{g}}_{k,p,j}}{\bf{g}}_{k,p,j}^H} } }, \\
		{\bm\nu } =& \sum\limits_{k = 1}^K {\sum\limits_{p = 1}^P {\sqrt {{\mu _{k,p}}}{{\bf{g}}_{k,p,k}}} }  \!-\! \sum\limits_{k = 1}^K {\sum\limits_{p = 1}^P {\sum\limits_{j = 1}^K {c_{k,p,j}^*{{\bf{g}}_{k,p,j}}} } }, \\
		{\zeta}  =&  \sum\limits_{k = 1}^K {\sum\limits_{p = 1}^P {\sum\limits_{j = 1}^K {{{\left| {{c_{k,p,j}}} \right|}^2}} } } 
		+ \sum\limits_{k = 1}^K {\sum\limits_{p = 1}^P {{\bm\varpi} _{k,p}^H{{\bf{\Xi }}_{k,p}}{{\bm\varpi} _{k,p}}} } 
		\\&-2\sum\limits_{k = 1}^K {\sum\limits_{p = 1}^P {\sqrt {{\mu _{k,p}}}}~{\mathfrak{R}}\left\{{c_{k,p,k}} \right\}}. \notag
	\end{align}
\end{subequations}
Therefore, the reformulated passive precoding subproblem ${\bar{\cal P}_{ \rm passive}}$ in (\ref{eqn:29}) can be further simplified as
\begin{equation}
	\label{eqn:37}
	\begin{aligned}
		{\hat{\cal P}_{ \rm passive}:}~~&\mathop {{\rm{min}}}\limits_{\bf{\Theta}}~~{{g}_6}({\bm{\Theta}}) =  {{\bm{\theta}}^H}{\bf{	\Lambda}}{\bm \theta} -{\mathfrak{R}} \left\{2 {{\bm{\theta}}^H}{\bm{\nu}}\right\} \\
		&~~{\rm{s.t.}}~~~C_2:{{\theta _{r,n}}}\in{\cal F},~\forall r\in{{\cal R}},\forall n\in{{\cal N}}.
	\end{aligned}
\end{equation}  
This simplified subproblem ${\hat{\cal P}_{ \rm passive}}$ is similar to those in \cite{Guo'19,Pan'19}. Since the matrix ${\bm\Lambda }$ is positive semidefinite, the objective function is convex. Besides, since we have ${{\cal F}} \triangleq \{ {\theta _{r,j}}{\Big |}\left| {{\theta _{r,j}}} \right| \le 1\}$ according to (\ref{eqn:4}), the constraint $C_2$ is also convex. Thereby, this subproblem ${\hat{\cal P}_{ \rm passive}}$ can be solved by ADMM \cite{admm}. 

However, similar to the optimization of active precoding vector $\bf W$, note that the adoption of ADMM \cite{admm} in this QCQP problem requires the matrix inversion for $\bm\Lambda$ with high computational complexity of ${\cal O}\left(R^3N^3\right)$. Since the RIS element number $N$ is usually very large, the complexity of using ADMM is very high. Again, to reduce the complexity, here we also provide a potential method based on the PDS to obtain the optimal solution ${\bf \Theta}^{\rm opt}$, which will be detailedly introduced in Appendix \ref{appendix:2}.

\section{Two-Timescale Extension of The Proposed Joint Precoding Framework}\label{sec:Ex}
Up to now, we have provided a joint precoding framework for the proposed RIS-aided cell-free network under the assumption of fully-known CSI in every small timescale, as shown in Fig. \ref{img:structure1}. However, due to the inherent high-dimensional channels introduced by RISs, acquiring all RIS-aided channels so frequently is usually unrealistic \cite{Huchen}. In particular, this issue is greatly exacerbated by the high-density of users, BSs, and RISs in RIS-aided cell-free networks, which prevents many existing joint precoding schemes based on fully-known CSI from practical adopting. 
\par
To tackle this issue, in this section, the proposed joint precoding framework is further extended to a more practical two-timescale scheme, which can serve as a trade-off scheme between overhead and performance. The key idea is to match each user with several well-performed RISs at the beginning of a large timescale. Then, in later several small timescales, only the RIS-aided channels of the matched user-RIS pairs are acquired and utilized for joint precoding design, while those of the unmatched pairs are temporarily ignored to relieve the pressure of RIS channel acquisitions.
\par
Specifically, this section is summarized as follows. Firstly, in Subsection \ref{subsec:Ex1}, the two-timescale extension of the proposed framework is overviewed. Then, the user-RIS matching problem is formulated in Subsection \ref{subsec:Ex2}. Finally, a LCR-based method is proposed to solve this problem in Subsection \ref{subsec:Ex3}.
\subsection{Overview of the proposed two-timescale extension}\label{subsec:Ex1}
In this subsection, the two-timescale extension of the proposed joint precoding framework is proposed as a more practical scheme to address the challenge of high-dimensional channels introduced by RISs in cell-free networks. 
\par
\begin{figure}[!t]
	\centering
	\includegraphics[width=3.4in]{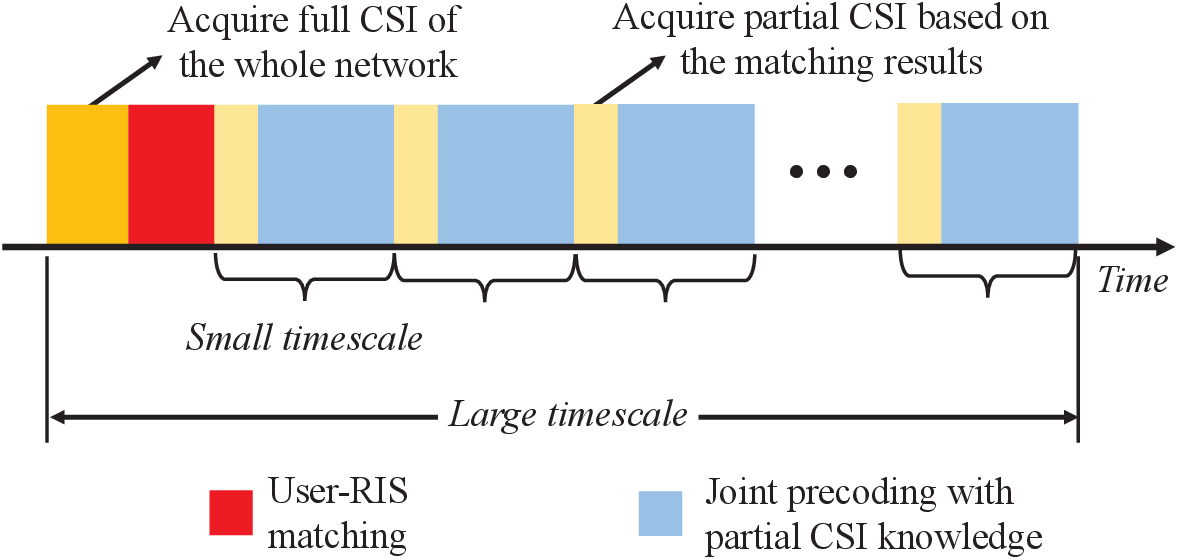}
	\caption{The dynamical working process of the two-timescale extension of the proposed joint precoding framework.}
	\label{img:structure}
	\vspace{-0.8em}
\end{figure}
Intuitively, we present how this two-timescale scheme works in Fig. \ref{img:structure}, which illustrates its dynamical working process over time in a large timescale. Particularly, the proposed two-timescale scheme exploits the properties that RISs far from users have little contribution to system capacity and the user's mobile scope is limited in a large timescale, thus the RISs far from users can be ignored temporarily. Specifically, at the beginning of a large timescale, all BS-RIS-user channels are fully acquired once. Then, under the principle of matching relatively strong user-RIS pairs, each user is matched with several well-performed RISs. Subsequently, in the next several small timescales, only the RIS-aided channels of the matched user-RIS pairs and the direct-link channels are acquired and utilized for precoding design while those unmatched are ignored. Finally, the above process will be repeated in the next large timescale.
\par
With this two-timescale scheme, only the limited CSI is acquired and utilized in small timescales, and it is unnecessary to acquire all RIS-aided channels so frequently as before. As the cost of the incomplete CSI, the two-timescale scheme will lead to performance loss, while we will show this loss is actually very limited in later Section \ref{sec:NSR}. Consequently, compared with the original scheme in Fig. \ref{img:structure1}, the two-timescale scheme in Fig. \ref{img:structure} is more practical to be applied, especially for the networks with a large number of distributed RISs. 
\subsection{User-RIS matching: Problem formulation}\label{subsec:Ex2}
To determine which RIS-aided channels should be acquired and utilized in small timescales as shown in Fig. \ref{img:structure}, in this subsection, the user-RIS matching problem is formulated.
\par
Firstly, let $u_{k,r}\in\{0,1\}$ denote the indicator variable that indicates whether the user $k$ and the RIS $r$ is matched. Particularly, $u_{k,r}$ being one/zero means that the channels between user $k$ and RIS $r$ will/won't be acquired and utilized in later small timescales. To simplify the expression, we define ${\bf u}_{k}=\left[{u}_{k,1},\cdots,{u}_{k,R}\right]^T$ and ${\bf u}= \left[{\bf u}_{1}^T,\cdots,{\bf u}_{K}^T\right]^T$. 
To determine which user-RIS pairs should be matched, we define the following virtual sum-rate ${\hat R}_{\rm sum}$ as the bonus function\footnote{ This bonus function is chosen for the expected result that, by maximizing ${\hat R}_{\rm sum}$, the user-RIS pairs with stronger links are more preferred to be matched, while those unmatched pairs have relatively little contribution to capacity improvement so that they can be temporarily ignored in later small timescales to relieve the pressure of RIS channel acquisitions.
} of the user-RIS matching problem:
\begin{align}
{\hat R}_{\rm sum}=\sum\limits_{k = 1}^K {\sum\limits_{p = 1}^P {{\eta_k}{{\log }_2}\left( {1 + {{\hat \gamma}_{k,p}}} \right)} }
\end{align}
where the virtual \ac{sinr} ${\hat \gamma}_{k,p}$ is equivalent to the real \ac{sinr} ${\gamma}_{k,p}$ in (\ref{eqn:8}) in which the end-to-end channels ${{\bf{h}}_{b,k,p}^H}$ are replaced by
\begin{equation}
	{{ \bf{\hat h}}_{b,k,p}^H} = {{\bf{H}}_{b,k,p}^H} + {\sum\limits_{r = 1}^R {{{{u}_{k,r}}{\bf{F}}_{r,k,p}^H{\bf{\Theta }}_r^H{{\bf{G}}_{b,r,p}}}} }.
\end{equation}
\par
Next, let $R_{\rm match}$ denote the maximum number of RISs that each user can be matched with ($R_{\rm match} \le R$), thus the user-RIS matching problem ${\cal P}_{\rm match}$ can be formulated as
\begin{align}\label{eqn:match}
	\!\!\!\!\!\!\!{\cal P}_{\rm match}:&\mathop {{\rm{max}}}\limits_{{\bf{u }},{\bf{W }},{\bm{\Theta }}} ~{\hat R}_{\rm sum}({\bf{u }},{\bf{W }},{\bm{\Theta }})\!=\!\sum\limits_{k = 1}^K {\sum\limits_{p = 1}^P {{\eta_k}{{\log }_2}\left( {1 + {{\hat \gamma}_{k,p}}} \right)} } \notag\\
	&~~~{\rm{s.t.}}~~\,C_1:\sum\limits_{k = 1}^K {\sum\limits_{p = 1}^P {\left\| {{{\bf{w}}_{b,p,k}}} \right\|}^2 }  \le {P_{b,\max }},~\forall b \in \mathcal{B}, \notag \\
	&~~~~~~~~~\,C_2:{{\theta _{r,n}}}\in{\cal F},~~~~~~~~~~\forall r\in{{\cal R}},\forall n\in{{\cal N}}, \\
	&~~~~~~~~~\,C_3: u_{k,r}\in\{0,1\},~~~~~\forall k\in{\cal K}, \forall r\in{\cal R}, \notag \\
	&~~~~~~~~~\,C_4: \sum\limits_{r = 1}^R {{u_{k,r}}}  \le R_{\rm match},~~\forall k\in{\cal K}, \notag
\end{align}
where $C_3$ is zero-one constraint of the indicator variable $\bf u$, and $C_4$ ensures that each user can only be matched with at most $R_{\rm match}$ RISs. 
\par
Our goal is to find a best-possible matching vector ${\bf u}^{\rm opt}$ to maximize the virtual sum-rate\footnote{ The virtual WSR ${\hat R}_{\rm sum}$ is just the bonus function for user-RIS matching instead of the real WSR, while the real WSR can be evaluated by substituting the optimized ${\bf W}^{\rm opt}$ and ${\bm \Theta}^{\rm opt}$ into $R_{\rm sum}$ in (\ref{eqn:9}).} ${\hat R}_{\rm sum}$ in user-RIS matching problem ${\cal P}_{\rm match}$ in (\ref{eqn:match}). However, due to the non-convexity of zero-one constraint $C_3$, the problem ${\cal P}_{\rm match}$ is challenging. Again, we consider to solve the problem via alternative optimization, which will be detailedly introduced in the next subsection.
\subsection{User-RIS matching: LCR-based method}\label{subsec:Ex3}
\begin{algorithm}[!t] 
	\caption{Proposed user-RIS matching method.} 
	\label{alg:2} 
	\begin{algorithmic}[1] 
		\REQUIRE ~~ 
	All channels ${{\bf{H}}_{b,k,p}},{\bf{G}}_{b,r,p}$ and $ {\bf{F}}_{r,k,p}$ where $\forall b\in {\cal B},k\in {\cal K},p\in {\cal P}$.
		\ENSURE ~~ 
		Optimized user-RIS matching vector $\bf{u}$.
		\STATE Initialize $\bm{\Theta}$, $\bm{W}$, and $\bf u$;
		\WHILE {no convergence of ${\hat R}_{\rm sum}$}
		\STATE Update ${\bm \rho}$ by (\ref{eqn:16});
		\STATE Update ${\bm \xi}$ by (\ref{eqn:20});
		\STATE Update ${\bf W}$ by solving (\ref{eqn:25});
		\STATE Update ${\bm \varpi}$ by (\ref{eqn:31});
		\STATE Update ${\bm{\Theta}}$ by solving (\ref{eqn:37});
		\STATE Update ${\bf{u}}$ by solving (\ref{eqn:45});
		\ENDWHILE	
		\RETURN ${\bf{u}}^{\rm opt}$. 
	\end{algorithmic}
\end{algorithm}
We consider to solve the matching problem ${\cal P}^{\rm m}$ in (\ref{eqn:match}) by optimizing ${\bf u}$, ${\bf W}$, and ${\bm \Theta}$ alternately. Note that, when ${\bf u}$ is fixed, the problem  ${\cal P}_{\rm match}$ in (\ref{eqn:match}) has the same form as the original problem ${\cal P}^{\rm o}$ in (\ref{eqn:11}). Therefore, the desired algorithm to solve ${\cal P}_{\rm match}$ can be realized by inserting a new step of optimizing $\bf u$ into {\bf Algorithm 1}, and here we summarize it as {\bf Algorithm 2} for clarity. In the rest of this subsection, we will focus on the step 8, i.e., optimizing ${\bf u}$ while fixing the other variables, to complete this algorithm.
\par
Firstly, we consider to reformulate ${\cal P}_{\rm match}$ as a solvable convex problem. Similar to the optimization of $\bm \Theta$, by applying LDR in {\it Proposition 1} and MCQT in {\it Proposition 3} and then fixing all auxiliary variables, the user-RIS matching problem ${\cal P}_{\rm match}$ can be reformulated as:
\begin{align}\label{eqn:38}
	{{\bar {\cal P}}_{ \rm match}:}~
	&\mathop {{\rm{max}}}\limits_{\bf{u}}~~{\hat{g}}\left({\bf{u }}\right)=\sum\limits_{k = 1}^K {\sum\limits_{p = 1}^P {{\hat g_{k,p}}\left( {\bf{u}} \right)} } \notag\\
		&~~{\rm{s.t.}}~~C_3: u_{k,r}\in\{0,1\},~~\forall k\in{\cal K}, \forall r\in{\cal R}, \\
	&~~~~~~~~C_4: \sum\limits_{r = 1}^R {{u_{k,r}}}  \le R_{\rm match},~~~\forall k\in{\cal K}, \notag
\end{align}
where  ${{\hat g}_{k,p}}({\bf{u }})$ is defined as
\begin{equation}
	\begin{aligned}
		&{{\hat g}_{k,p}}({\bf{u }}) = 2\sqrt {{\mu}_{k,p}}\, {\mathfrak{R}} \left\{ {{\bm{\varpi }}_{k,p}^H{{\bf{\hat Q}}_{k,p,k}}\left( {\bf{u }} \right)} \right\}-\\&~~~~~~~ {\bm{\varpi }}_{k,p}^H\left( \sum\limits_{j = 1}^K {{{\bf{\hat Q}}_{k,p,j}}\left( {\bf{u }} \right){\bf{\hat Q}}_{k,p,j}^H\left( {\bf{u }} \right)}  + {{\bf{\Xi }}_{k,p}} \right){{\bm{\varpi }}_{k,p}},
	\end{aligned}
\end{equation} 
in which the auxiliary functions ${{\bf{\hat Q}}_{k,p,j}}\left({\bf u}\right)\in{\mathbb C}^{U}$ satisfy
\begin{equation}
	{{\bf{\hat Q}}_{k,p,j}}\left( {\bf{u }} \right) \!=\! \sum\limits_{b = 1}^B \! {\left({{\bf{H}}_{b,k,p}^H} + {\sum\limits_{r = 1}^R {{{{u}_{k,r}}{\bf{F}}_{r,k,p}^H{\bf{\Theta }}_r^H{{\bf{G}}_{b,r,p}}}} } \right){{\bf{w}}_{b,p,j}}}.
\end{equation}

Then, we wish to simplify the expression of the objective function ${\hat g}$ by exploiting the following definitions:
\begin{subequations}\label{eqn:42m}
\begin{align}
	{\beta _{k,p,j,r}}=\sum\limits_{b = 1}^B {\bm \varpi _{k,p}^H{\bf{F}}_{r,k,p}^H{\bf{\Theta }}_r^H{{\bf{G}}_{b,r,p}}{{\bf{w}}_{b,p,j}}},
	\\
	{{\bm{\beta }}_{k,p,j}} = \left[ {{\beta _{k,p,j,1}},{\beta _{k,p,j,2}}, \cdots ,{\beta _{k,p,j,R}}} \right]^T.
\end{align}
\end{subequations}
After substituting (\ref{eqn:c_kpj}) and (\ref{eqn:42m}) into the objective function ${\hat{g}}$ in (\ref{eqn:38}) and then removing the terms unrelated to ${\bf u}$, ${\hat{g}}$ in (\ref{eqn:38}) can be rewritten as
\begin{align}
{\hat {g}}_1\left({\bf{u }}\right)=-{{\bf{u}}^T}{\bf{\Omega u}} + 2{\bm{\zeta }}^T{{\bf{u}}}	
\end{align}
where
\begin{align}
	{\bf{\Omega }} \!=\! {\rm{diag}}\!\left( {{{\bf{\Omega }}_1},{{\bf{\Omega }}_2}, \cdots ,{{\bf{\Omega }}_K}} \right),~
	{\bm{\zeta }} \!=\! {\left[ {{\bm{\zeta }}_1^T,{\bm{\zeta }}_2^T, \cdots ,{\bm{\zeta }}_K^T} \right]^T},
\end{align}
in which ${{\bf{\Omega }}_k}$ and ${{\bm{\zeta }}_k}$ are given by
\begin{subequations}
	\begin{align}
		{{\bf{\Omega }}_k} &= \Re \left\{ \sum\limits_{p = 1}^P {\sum\limits_{j = 1}^K {{{\bm{\beta }}_{k,p,j}}{\bm{\beta }}_{k,p,j}^H} }\right\},\\
		{{\bm{\zeta }}_k} &= \Re \left\{ {\sum\limits_{p = 1}^P {\sqrt {{\mu _{k,p}}} {{\bm{\beta }}_{k,p,k}}}  - \sum\limits_{p = 1}^P {\sum\limits_{j = 1}^K {c_{k,p,j}^*{{\bm{\beta }}_{k,p,j}}} } } \right\}.
	\end{align}
\end{subequations}
Thus, ${\bar {\cal P}}_{ \rm match}$ in (\ref{eqn:38}) can be equivalently reorganized as
\begin{align}\label{eqn:45}
	{{\cal {\hat P}}_{ \rm match}:~}
	&\mathop {{\rm{min}}}\limits_{\bf{u}}~{\hat {g}}_2\left({\bf{u }}\right)={{\bf{u}}^T}{\bf{\Omega u}} - 2{\bm{\zeta }}^T{{\bf{u}}}	
	 \notag\\
	&~~{\rm{s.t.}}~C_3: \left[ {\bf{u}} \right]_j^2 = {\left[ {\bf{u}} \right]_j} ,~\forall j \in \left\{ {1, \cdots ,KR} \right\}, \\
	&~~~~~~~C_4: \left( {{{\bf{I}}_K} \otimes {\bf{1}}_R^T} \right){\bf{u}} \le {R_{\rm match}}{{\bf{1}}_K}, \notag
\end{align}
where the constraints $C_3$ and $C_4$ are also equivalently rewritten for clearer expression.
\par
Note that, the equivalent problem ${\cal {\hat P}}_{ \rm match}$ in (\ref{eqn:45}) is actually a zero-one quadratic programming problem with linear constrains, which is proved to be NP-hard \cite{NP-hard}. Since the dimension of the matching vector ${\bf{u}}$ is $KR$, when the number of users $K$ and RISs $R$ is not too large, Bruto-Force method can be directly used to search for the optimal solution to ${\cal {\hat P}}_{ \rm match}$. However, when the densities of users and RISs in cell-free network are high, it is unpractical to adopt Bruto-Force due to the large search space ($2^{KR}$ candidates). In this case, we can apply linear conic relaxation (LCR) \cite{LCR} to relax the zero-one programming problem ${\cal {\hat P}}_{ \rm match}$ and try to find a sub-optimal solution as a substitute. In this way, we obtain the following proposition.
\par
\begin{proposition}\label{proposition:4}
Applying LCR \cite{LCR} for the zero-one programming problem ${\cal {\hat P}}_{ \rm match}$ in (\ref{eqn:45}) and by introducing an symmetric auxiliary matrix ${\bf U}\in{\mathbb R}^{KR\times KR}$ to replace the matrix ${\bf{u}}{{\bf{u}}^T}$, ${\cal {\hat P}}_{ \rm match}$ can be reformulated as
\begin{align}\label{eqn:46}
	{{\cal {\tilde P}}_{ \rm match}\!:}~
	&\mathop {{\rm{min}}}\limits_{{\bf{u}},{\bf{U}}}~{\hat {g}}_2\left({\bf{u}},{\bf U} \right)={\rm{Tr}}\!\left( {{\bf{\Omega U}}} \right) - 2{{\bm{\zeta }}^T}{\bf{u}}
	\notag\\
	&~~{\rm{s.t.}}~C_3: {\left[ {\bf{U}} \right]_{j,j} \!=\! {\left[ {\bf{u}} \right]_j}},~~\forall j \in \left\{ {1, \cdots ,KR} \right\}, \notag \\
	&~~~~~~~C_4: \left( {{{\bf{I}}_K} \otimes {\bf{1}}_R^T} \right){\bf{u}} \le {R_{\rm match}}{{\bf{1}}_K},\\
	&~~~~~~~C_5: {\left[ {\bf{U}} \right]_{i,j} \!=\! {\left[ {\bf{U}} \right]_{j,i}}},~~\forall i,j \in \left\{ {1, \cdots ,KR} \right\}, \notag	
	 \\
	&~~~~~~~C_6: \left( {{{\bf{I}}_K} \otimes {\bf{1}}_R^T} \right){\bf{U}} \le {R_{\rm match}}{{\bf{1}}_K}{{\bf{u}}^T}, \notag\\	
	&~~~~~~~C_7: \left[ {\begin{array}{*{20}{c}}
			1&{{{\bf{u}}^T}}\\
			{\bf{u}}&{\bf{U}}
	\end{array}} \right]\succcurlyeq 0, \notag
\end{align}
where the newly added constraints $C_5$, $C_6$, $C_7$ are the guarantees for ${\bf U}$ to approach ${\bf{u}}{{\bf{u}}^T}$ as close as possible, and the rank-one constraint of ${\bf U}$ is relaxed.
\end{proposition}

Notice that, the constraint $C_3$ in the problem ${{\cal {\tilde P}}_{ \rm match}}$ in (\ref{eqn:46}) is now convex, and thus ${{\cal {\tilde P}}_{ \rm match}}$ is actually a solvable semidefinite programming (SDP) problem. Thereby, the optimal solution ${\bf u}^{\rm opt}$ to ${{\cal {\tilde P}}_{ \rm match}}$ can be efficiently obtained by adopting the modern SDP solvers such as SDPT3 \cite{SDPT3}, of which the computational complexity is about ${\cal O}\left(K^2R^2\left(KR+1\right)^{2.5}\right)$ in the worst case \cite{Yang'14}.

\section{Framework Supplements}\label{sec:FS}
In this section, more supplements of the proposed joint precoding framework are provided. Specifically, limited by the practical hardware implementation, we discuss the non-ideal RIS case in Subsection \ref{sec:Alg:p5}. Then, we analyze the convergency and computational complexity of the proposed algorithms in Subsection \ref{sec:Alg:p6} and  Subsection \ref{sec:Alg:p7}, respectively.
\subsection{Extension to non-ideal RIS cases}\label{sec:Alg:p5}
Up to now, we have provided a complete joint precoding framework for the proposed RIS-aided cell-free network, where ideal RIS case mentioned in Subsection \ref{channel} is considered. In this subsection, we will extend this framework to the more practical non-ideal \ac{ris} cases.
\subsubsection{Non-ideal \ac{ris} cases}
According to (\ref{eqn:4}) in Subsection \ref{channel}, ${\cal F}$ is defined as the ideal RIS case, where both the amplitude and phase of ${\theta _{r,n}}$ associated with the \ac{ris} element can be controlled independently and continuously. However, limited by the hardware implementation of metamaterials, the \ac{ris}s in practice are usually non-ideal. For consistent discussion, here we redefine the ideal \ac{ris} case ${\cal F}$ as ${\cal F}_1$. We also define ${\cal F}_2$ as the general case where only the phase of ${\theta _{r,j}}$ can be controlled continuously \cite{Wu'19}, and ${\cal F}_3$ as the practical case where the low-resolution phase of ${\theta _{r,n}}$ is discrete \cite{LinglongDai}, i.e., 
\begin{subequations}
	\begin{align}
		&{{\cal F}_1} \triangleq \left\{ {\theta _{r,n}}{\Big |} \left| {{\theta _{r,n}}} \right| \le 1\right\},\\
		&{{\cal F}_2} \triangleq \{ {\theta _{r,n}}{\Big |}\left| {{\theta _{r,n}}} \right| = 1\},\\
		&{{\cal F}_3} \triangleq \left\{ {\theta _{r,n}}{\Big |}{\theta _{r,n}} \in \left\{ 1,{e^{j\frac{{2\pi }}{L}}}, \cdots ,{e^{j\frac{{2\pi (L - 1)}}{L}}}\right\} \right\},
	\end{align}
\end{subequations}
where $L$ indicates that ${{\cal F}_3}$ contains $L$ discrete phase shifts.
\par
According to the simplified subproblem $\hat{\cal P}_{\rm passive}$ in (\ref{eqn:37}), for the passive precoding design at the RISs, when the phase shift constraint ${\cal F}$ is the ideal case ${\cal F}_1$ as discussed above, the subproblem ${\hat{\cal P}_{ \rm passive}}$ can be directly solved due to the convexity of ${\cal F}_1$. However, when the constraint ${\cal F}$ in (\ref{eqn:37}) is non-ideal ${\cal F}_2$ or ${\cal F}_3$, the simplified subproblem ${\hat{\cal P}_{ \rm passive}}$ in (\ref{eqn:37}) becomes non-convex. To address such problems, we provide the feasible solutions in the following two parts, respectively.

\subsubsection{Optimal solution in continuous phase shift case ${\cal F}_2$}
When ${\cal F}={\cal F}_2$, the subproblem (\ref{eqn:37}) can be equivalently rewritten as
\begin{equation}
	\label{eqn:con}
	\begin{aligned}
		{\hat{\cal P}_{\text{passive-con}}:}~~&\mathop {{\rm{min}}}\limits_{\bf{\Theta}}~~{{g}_6}({\bm{\Theta}}) =  {{\bm{\theta}}^H}{\bf{	\Lambda}}{\bm \theta} -{\mathfrak{R}} \left\{2 {{\bm{\theta}}^H}{\bm{\nu}}\right\}\\
		&\,{\rm{s.t.}}~~~C_2:\left| {{\theta _{r,n}}} \right|{\rm{ = 1}},~\forall r\in{{\cal R}},\forall n\in{{\cal N}},
	\end{aligned}
\end{equation}  
which is non-convex due to the constant modulus constraint. Fortunately, a similar subproblem has been discussed in \cite[Eqn. 41]{Pan'19}, while it has been proved that its optimal solution can be obtained by adopting the majorization-minimization (MM) algorithm \cite{Sun'17} or complex circle manifold (CCM) algorithm \cite{Alhujaili'19}.
\subsubsection{Approximation in discrete phase shift case ${\cal F}_3$}
When ${\cal F}={\cal F}_3$, the common solution to address the non-convex constraint of discrete space is approximation projection \cite{Wu'19'2,Nadeem'19,Li'19}. Specifically, we can first relax ${\cal F}_3$ to ${\cal F}_2$, and obtain the optimal solution ${\bm\theta}^{\rm opt}$ by solving (\ref{eqn:con}). Then, following the proximity principle, we can simply project the solved ${\bm\theta}^{\rm opt}$ to the elements in ${\cal F}_3$ by a approximation projection, written as
\begin{equation}
	\label{eqn:39}
	\angle \theta _{r,j}^{\rm sub}{\rm{ = }}\mathop {{\rm{argmin}}}\limits_{\phi  \in {\cal F}_3} \left| {\angle \theta _{r,j}^{\rm opt} - \angle \phi  } \right|,~~\forall r\in{{\cal R}},\forall j\in{{\cal N}},
\end{equation}
where $\theta _{r,j}^{\rm sub}$ denotes the approximated sub-optimal solution to the phase shift $\theta _{r,j}$. Thus, we can finally obtain the sub-optimal solution ${\bm {\Theta} }^{\rm sub}\in{\cal F}_3$ to the subproblem (\ref{eqn:37}).
\subsection{Algorithm convergency}\label{sec:Alg:p6}
In the ideal case ${\cal F}={\cal F}_1$ and general case ${\cal F}={\cal F}_2$, the proposed joint precoding framework, has strict convergency\footnote{ Note that, the proposed joint precoding scheme converges to a feasible solution, while only the convergence rate is fully analyzed can its local optimality be strictly proved.}, since each step of the iteration, i.e., (\ref{eqn:16}), (\ref{eqn:20}), (\ref{eqn:25}), (\ref{eqn:31}) and (\ref{eqn:37}), can be easily proved to be monotonous. However, in the non-ideal case ${\cal F}={\cal F}_3$, the convergency of the proposed framework can not be proved strictly, since the update of ${\bm{\Theta}}$ has no guarantee of monotony in some cases due to the approximation operation in (\ref{eqn:39}). Besides, for {\bf Algorithm 2}, how the relaxation operation of optimizing the user-RIS matching vector $\bf u$ will influence the global convergency is also unpredictable \cite{LCR}. Fortunately, since the rest of iterative steps are all monotonous, actually the loss caused by the approximation operation and relaxation operation have little adverse effect on the global convergence, which will be verified through simulations in the next Section \ref{sec:NSR}. 
\subsection{Computational complexity}\label{sec:Alg:p7}
\begin{table}[t]
	\centering
	\normalsize
		\caption{Computational complexity for updating each variable.}
		\label{table:1}
		\begin{tabular}{|c|c|c|c|c|c|c|}
			\toprule  
			Variable&Computational complexity ${\cal O}\left(\cdot\right)$\\ 
			\midrule  
			${\bm \rho}$& $PK\left(KUBM+U^3+(K+1)U^2+U\right)$ \\ \hline
			${\bm \xi }$&$PK\left(KUBM+U^3+(K+1)U^2\right)$ \\ \hline
			${\bf W}$& ${{I_{\rm{a}}}\left( {{B^2}{M^2}{P^2}{K^2} + BMPK} \right)}$ \\ \hline
			${\bm \varpi }$& $PK\left(U^3+(K+1)U^2\right)$\\ \hline
			${\bm {\Theta} }$&${{I_{\rm{p}}}\left( {{R^2}{N^2} + RN} \right)}$\\ \hline
			$ {\bf {u} }$&$ K^2R^2\left(KR+1\right)^{2.5}$	\\		
			\bottomrule  
		\end{tabular}
		\vspace*{-1em}
\end{table}
The overall computational complexities of the proposed joint precoding framework  and its two-timescale extension are mainly introduced by the updates of the variables ${\bm \rho}$, ${\bm \xi }$, ${\bf W}$, ${\bm \varpi }$, ${\bm {\Theta} }$, and ${\bf {u} }$, as shown in (\ref{eqn:16}), (\ref{eqn:20}), Appendix \ref{appendix:1}, (\ref{eqn:31}), Appendix \ref{appendix:2}, and (\ref{eqn:46}), respectively. To make it clear, we summarize their computational complexities in Tab. \ref{table:1}, wherein $I_{\rm a}$ and $I_{\rm p}$ are the required iteration numbers for the convergence of the subproblems ${\hat{\cal P}_{ \rm active}}$ in (\ref{eqn:25}) and ${\hat{\cal P}_{ \rm passive}}$ in (\ref{eqn:37}), respectively.

From Tab. \ref{table:1}, we can observe that the updates of ${\bm \rho}$, ${\bm \xi }$, and ${\bm \varpi }$ have much lower computational complexity than ${\bf W}$, ${\bm {\Theta} }$, and ${\bf u}$, since ${\bm \rho}$, ${\bm \xi }$, and ${\bm \varpi }$ have close-form update formula while ${\bf W}^{\rm opt}$, ${\bm {\Theta} }^{\rm opt}$, and ${\bf u}^{\rm opt}$ can only be obtained by solving QCQP and SDP problems via iterative optimization methods. Furthermore, since the density of BSs and users in cell-free network is high \cite{Ngo'17} and the number of RIS elements is usually large \cite{Wang'19}, we can reasonably assume that $BMPK \gg U$ and $RN \gg U$. Under this assumption, the overall computational complexity of {\bf Algorithm 1} can be approximated by ${\cal O}\left(I_{\rm o}\left( {{I_{\rm{a}}}{B^2}{M^2}{P^2}{K^2} + {I_{\rm{p}}}{R^2}{N^2}} \right)\right)$ and that of {\bf Algorithm 2} can be similarly approximated by ${\cal O}\left(I_{\rm o}\left( {{I_{\rm{a}}}{B^2}{M^2}{P^2}{K^2} + {I_{\rm{p}}}{R^2}{N^2}}+K^2R^2\left(KR+1\right)^{2.5} \right)\right)$, wherein $I_{\rm o}$ denotes the required number of iterations for the algorithm convergence.

\section{Simulation Results}\label{sec:NSR}
In this section, we provided extensive simulation results under different conditions to validate the performance of the proposed concept of RIS-aided cell-free network. 
\subsection{Simulation setup}
\begin{figure}[!t]
	\centering
	\includegraphics[width=3.4in]{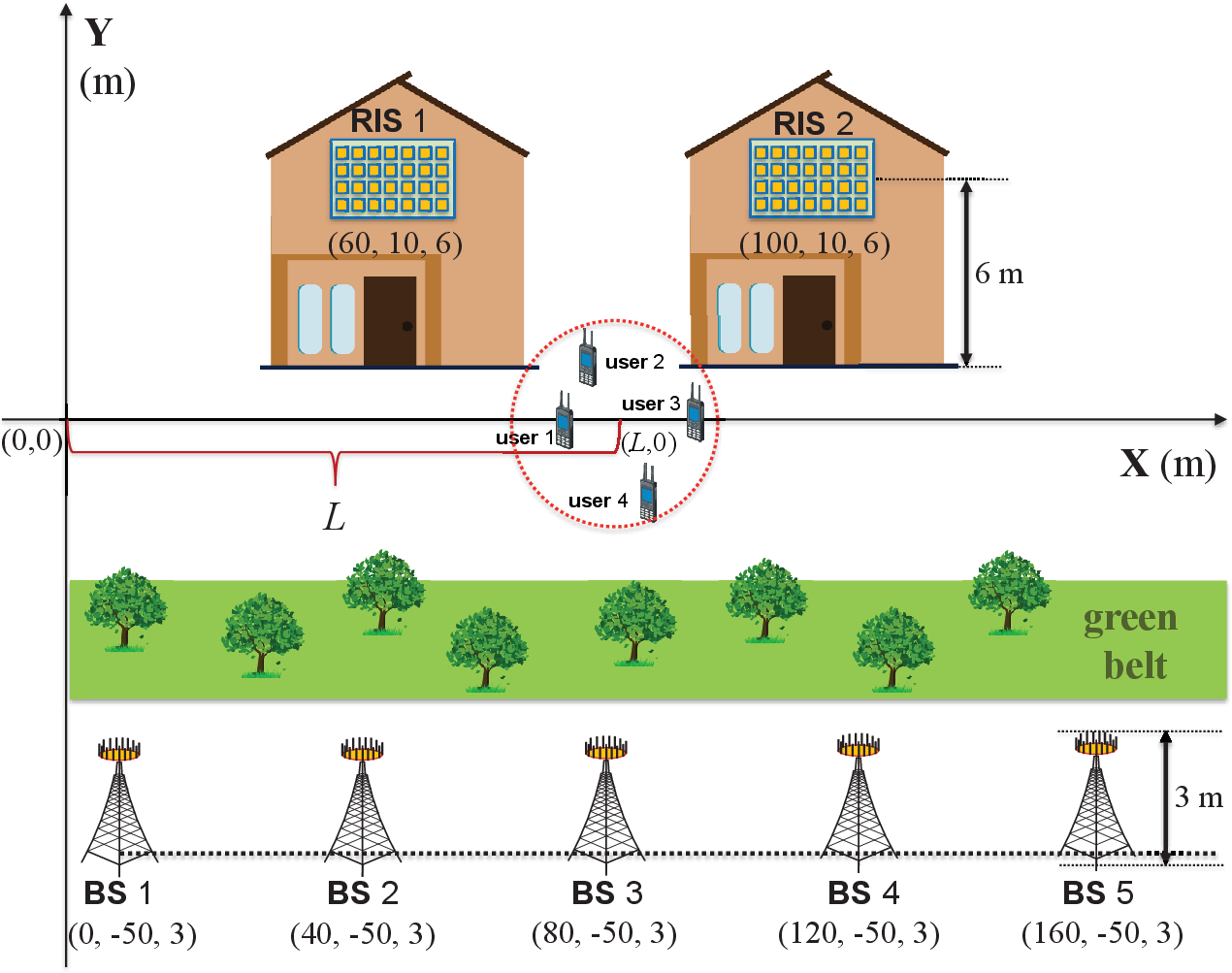}
	\caption{The simulation scenario where five \ac{bs}s assisted by two \ac{ris}s serve four users.}
	\label{img:simulation_scenario_1}
	\vspace{-0.8em}
\end{figure}

For the simulation of the proposed RIS-aided cell-free network, we draw on the potential deployment schemes for cell-free network from the existing works \cite{Interdonato'19} to construct simulation scenarios. For simplicity but without loss of generality, we consider a 3-D scenario with the topology shown in Fig. \ref{img:simulation_scenario_1}. In this setup, a cell-free network with five \ac{bs}s serve four users simultaneously, while the network capacity is limited due to the obstruction of a green belt. To improve the capacity, two \acp{ris} are separately deployed on two distant building surfaces, which are high enough to construct extra reflection links. We assume the $i$-th BS is located at $\left(40\left(i-1\right)\,{\text m},-50\,{\text m},3\,{\text m}\right)$, and two \acp{ris} are located at $(60\,{\text m},10\,{\text m},6\,{\text m})$ and $(100\,{\text m},10\,{\text m},6\,{\text m})$, respectively. Since the \ac{bs} in cell-free network is usually small with few antennas and low transmit power \cite{Interdonato'19}, we further assume the maximum transmit power at the \ac{bs} is set as $P_{b,\max}=0$ dBm, while the number of antennas at each \ac{bs} is set as $M=2$ and that at each user is set as $U=2$. The number of \ac{ris} elements is set as $N=100$. The number of subcarriers is set as $P=4$, and the noise power is set as ${\sigma ^2} =  - 80$ dBm \cite{Wu'19}.
\par
 For the channel model, we consider the same settings as those in \cite{Wu'19}. For the large-scale fading model, firstly we define $d_{\rm Bu}$, $d_{\rm BR}$, $d_{\rm Ru}$ as the distance between \ac{bs} and user, \ac{bs} and \ac{ris}, \ac{ris} and user, respectively. Thus the distance-dependent path loss model is given by
	\begin{equation}\label{eqn:40}
		{L}({d}) = {C_{0}}{\left(\frac{d}{d_0}\right)}^{ - {\kappa}},~~d\in\{d_{\rm Bu},d_{\rm BR},d_{\rm Ru}\}
	\end{equation}
	where $C_{0}$ is the path loss at the reference distance $d_0=1\,{\text m}$ and $\kappa$ denotes the path loss exponent. Here we assume $C_0=-30$ dB, and the path loss exponents of the BS-RIS link, RIS-user link, and BS-user link are set as ${\kappa _{\rm BR}}=2.2$, ${\kappa _{\rm Ru}}=2.8$, and ${\kappa _{\rm Bu}}=3.5$ \cite{Wu'19}, respectively. To account for the small-scale fading, we further consider a Rician fading channel model, thus the BS-user channel ${\bf H}$ is obtained by 
\begin{equation}
	\label{eqn:42}
	\begin{aligned}
		{\bf H}=\sqrt{\frac{\omega_{\mathrm{Bu}}}{1+\omega_{\mathrm{Bu}}}} {\bf H}^{\mathrm{LoS}}+\sqrt{\frac{1}{1+\omega_{\mathrm{Bu}}}} {\bf H}^{\mathrm{NLoS}},
	\end{aligned}
\end{equation}
where $\omega_{\mathrm{Bu}}$ denotes the Rician factor, and ${\bf H}^{\mathrm{LoS}}$ and ${\bf H}^{\mathrm{NLoS}}$ denote the LoS and Rayleigh fading components, respectively. Note that ${\bf H}$ is equivalent to a LoS channel when $\omega_{\mathrm{Bu}} \rightarrow \infty$, and a Rayleigh fading channel when $\omega_{\mathrm{Bu}} =0$. Then, ${\bf H}$ is multiplied by the square root of the distance-dependent path loss ${L}({d_{\rm Bu}})$ in (\ref{eqn:40}). Similarly, the BS-RIS and RIS-user channels can also be generated by the above procedure, and let $\omega_{\mathrm{BR}}$ and $\omega_{\mathrm{Ru}}$ denote the Rician factors of them, respectively. We further assume $\omega_{\mathrm{BR}}\rightarrow\infty$, $\omega_{\mathrm{Bu}}=0$ and, $\omega_{\mathrm{Ru}}=0$ \cite{Wu'19}.
\par
At last, for the joint precoding framework, we set the weights of users as $\eta_{k}=1$. As an alternating algorithm, $\bm \Theta$ is initialized by random values in ${\cal F}$, $\bf W$ is initialized by identical power and random phases, and $\bf u$ is initialized by setting all of its elements to one.
\subsection{Weighted sum-rate of the RIS-aided cell-free network}\label{subsec:EV}
\begin{figure}[!t]
	\centering
	\includegraphics[width=3.7in]{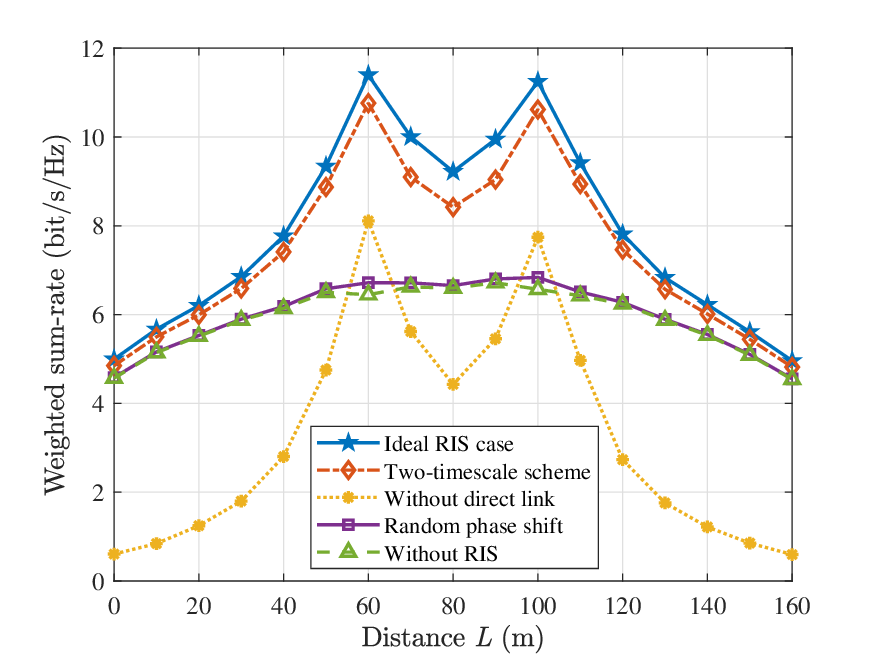}
	\caption{Weighted sum-rate against the distance $L$.}
	\label{img:simulation_1}
	\vspace{-0.8em}
\end{figure}
The \ac{wsr} of the proposed RIS-aided cell-free network is evaluated in this subsection. Firstly, we assume that four users are randomly distributed in a circle centered at $(L,0)$ with radius $\rm 1\,{\rm m}$. The height of these users is set as $1.5\,{\rm m}$, and we also assume the ideal RIS case (${\cal F}={\cal F}_1$). Then, we plot the \ac{wsr} against the distance $L$ in Fig. \ref{img:simulation_1}, in which the five curves are defined as follows:
	\begin{itemize}
		\item {\it Ideal RIS case:} Based on the fully-known CSI, {\bf Algorithm 1} is performed to maximize the WSR.	
	    \item {\it Two-timescale scheme:} This curve is realized in three steps. First, {\bf Algorithm 2} is performed to match each user with no more than $R_{\rm match}=1$ RISs. Then, only the RIS-aided channels of the matched user-RIS pairs and the direct-link channels are utilized for joint precoding design through {\bf Algorithm 1}. Finally, WSR is obtained by substituting the optimized ${\bf W}^{\rm opt}$ and ${\bm \Theta}^{\rm opt}$ from {\bf Algorithm 1} into $R_{\rm sum}$ in (\ref{eqn:9}).		
		\item {\it Without direct link:} Assume that all direct links between BSs and users are completely obstructed (${\bf H}=0$), and then {\bf Algorithm 1} is performed to maximize the WSR.	
		\item {\it Random phase shift:} All the phase shifts of RIS elements are randomly set to the values in ${\cal F}_1$. Then, based on the combined channels, {\bf Algorithm 1} is only performed at BSs to maximize the WSR.
		\item {\it Without RIS:} The conventional cell-free network without \ac{ris}. Based on the CSI of BS-user channels, the multi-user precoding method in \cite{Shen'18'1} is performed at BSs to maximize the WSR.
\end{itemize} 
\par
From Fig. \ref{img:simulation_1}, we have three observations. First, for the schemes with RIS deployed, we can see two obvious peaks at $L=60\,{\rm m}$ and $L=100\,{\rm m}$. It indicates that the \ac{wsr} rises when the users approach one of the two \ac{ris}s, since the users can receive strong signals reflected from the \ac{ris}s. While for the conventional scheme without \ac{ris}, these two peaks will not appear. Thus, we can conclude that the network capacity can be substantially increased by deploying RISs in the network, and the signal coverage can be accordingly extended. Second, we notice that the performance of “{\it Random phase shift}” has very limited gain compared with the scheme without RIS. The reason is that, without passive beamforming at RISs, the signals reaching RISs cannot be accurately directed to the users, which demonstrates the necessity of passive precoding. At last, we notice that, compared with “{\it Ideal RIS case}”, “{\it Two-timescale scheme}” suffers an averaged performance loss of about $10\%$. The reason is that, “{\it Ideal RIS case}” acquires and utilizes all RIS-aided channels ($KR=8$) for joint precoding design, while “{\it Two-timescale scheme}” only acquires and utilizes the RIS-aided channels of the matched user-RIS pairs ($KR_{\rm match}=4$). In later Subsection VI-E, we will analyze the system WSR with different $R_{\rm match}$ in detail.
\begin{figure}[!t]
	\centering
	\includegraphics[width=3.7in]{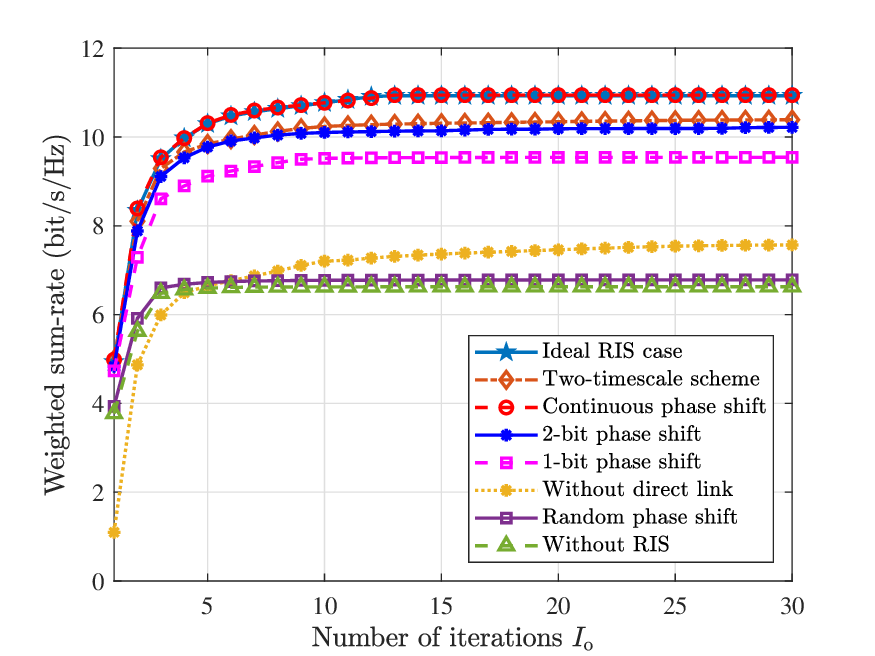}
	\caption{Weighted sum-rate against the number of iterations $I_{\rm o}$.}
	\label{img:simulation_2}
	\vspace{-0.8em}
\end{figure}
\subsection{Convergency of the joint precoding framework}\label{subsec:convergency}
To show the convergency of the proposed algorithms, we plot the \ac{wsr} against the number of iterations $I_{\rm o}$ in Fig. \ref{img:simulation_2} by running algorithms once with random initializations. The simulation setups are same as those used in Subsection \ref{subsec:EV}, and we fix the distance $L$ as $L=65\,{\rm m}$. To evaluate the convergency in non-ideal RIS phase shift cases, we add the curves “{\it Continuous phase shift}”, “{\it 1-bit phase shift}”, and “{\it 2-bit phase shift}” to denote the cases ${\cal F}_2$ and ${\cal F}_3$. The results in Fig. \ref{img:simulation_2} illustrate that, when the convergence error is no more than $1\%$, the proposed joint precoding framework can converge within 15 iterations. To be more specific, the ideal \ac{ris} case, two-timescale scheme, continuous phase shift case, and the case without direct links can converge within 15 iterations, while the discrete phase shift case can converge within 10 iterations. Since the conventional cell-free network without RIS and the scheme “{\it Random phase shift}” do not need to address the RIS precoding, the WSR in these cases can converge within 5 iterations. The results also indicate that, although the approximation operation in (\ref{eqn:39}) causes the uncertainty of convergence, the proposed framework still enjoys a fast global convergence.
More importantly, we find that the curves “{\it Ideal RIS case}” and “{\it Continuous phase shift}” are very close. The essential reason is that, for “Ideal RIS case”, actually most of the optimized RIS elements have an amplitude equal to 1. It indicates that, for the precoding design of the amplitude-uncontrollable RIS, we can simply relax its constant-modulus constraint and optimize its precoding as an ideal RIS. Then, we can directly project the optimized precoding matrix to the constant-modulus and obtain the final design. In this way, the optimization of RIS precoding becomes simple and low-complexity. Besides, it also implies that there is no need to design an amplitude-controllable \ac{ris} for wireless communications.

\subsection{Robustness of the joint precoding to CSI error}
	\begin{figure}[!t]
		\centering
		\includegraphics[width=3.7in]{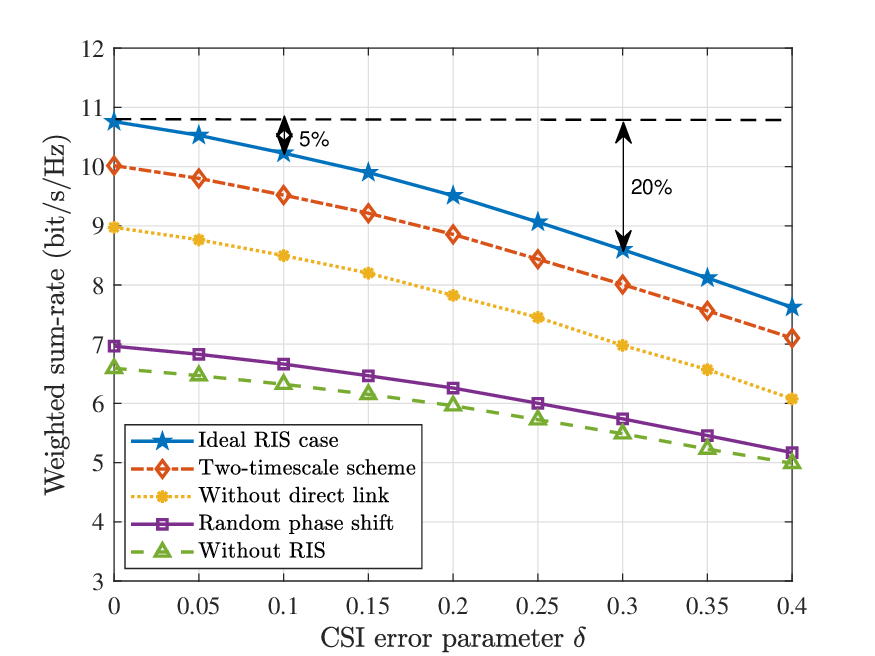}
		\caption{Weighted sum-rate against the CSI error parameter $\delta$.}
		\label{img:simulation_CSI_error}
		\vspace{-0.8em}
	\end{figure}
	Generally, in the proposed RIS-aided cell-free system, the channel estimation is very challenging due to the high-dimensional channels. Here we analyze the robustness of the proposed joint precoding scheme to CSI error. To generate the imperfect channels, we model the practically estimated channel $\hat h$ as \cite{Ubaidulla'11}
	\begin{equation}
		{\hat h}=h+e,
	\end{equation}
	where $h$ denotes the real channel and $e$ represents the estimation error with Gaussian distribution and zero mean, i.e. $e\sim{\cal CN}\left(0,\sigma^2_{e}\right)$. We assume the variance $\sigma^2_{e}$, i.e. the error power, satisfies $\sigma^2_{e}\triangleq\delta{\left| h \right|^2}$ wherein $\delta$ denotes the ratio of the error power $\sigma^2_{e}$ to the channel gain $\left| h \right|^2$, which characterizes the level of CSI error. Then, using the same setups in Subsection \ref{subsec:convergency}, we plot the WSR against the CSI error parameter $\delta$ in Fig. \ref{img:simulation_CSI_error}. From this figure, we can observe that, the performance loss grows with the increasing of $\delta$. Particularly, for the “{\it Ideal RIS case}”, compared with the perfect CSI without error (i.e. $\delta=0$), the system performance suffers a loss of $5\%$ when the error power is $10\%$ of the channel gain (i.e. $\delta=0.1$), and a loss of $20\%$ when $\delta=0.3$. Thereby, the proposed joint precoding scheme shows strong robustness to CSI error. 
\subsection{The impact of key system parameters}
To reveal more insights of the proposed RIS-aided cell-free network with different system parameters, we consider a new simulation scenario with more dispersed users and reset the locations of the four users in Fig. \ref{img:simulation_scenario_1} as $\left(L-60\,{\rm m},0\right)$, $\left(L-20\,{\rm m},0\right)$, $\left(L+20\,{\rm m},0\right)$, and $\left(L+60\,{\rm m},0\right)$, respectively.
\subsubsection{WSR against BS transmit power}
\begin{figure}[!t]
	\centering
	\includegraphics[width=3.7in]{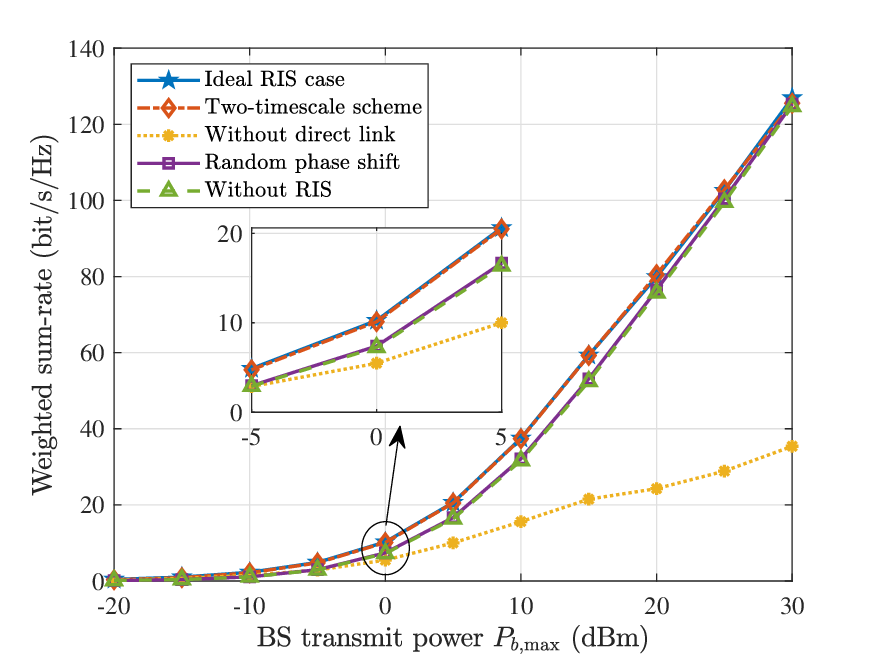}
	\caption{Weighted sum-rate against the \ac{bs} transmit power $P_{b,\max}$.}
	\label{img:simulation_3}
	\vspace{-0.8em}
\end{figure}
By fixing $L=80\,{\rm m}$, we present the average weighted sum-rate against the \ac{bs} transmit power in Fig. \ref{img:simulation_3}. From this figure we can observe that, with the increase of the \ac{bs} transmit power, the weighted sum-rate rises rapidly in all cases.  Particularly, “{\it Two-timescale scheme}” achieves very similar performance to “{\it Ideal RIS case}”. It indicates that, the RIS-aided channels of the unmatched user-RIS pairs indeed have negligible contribution to capacity improvement, which demonstrates the effectiveness of the proposed two-timescale joint precoding scheme. Besides, we have an important finding that, the performance gain brought by \ac{ris}s is significant only when the BS transmit power is moderate (e.g., form 0 dBm to 20 dBm), while the performance gain becomes negligible when the BS transmit power is too low (e.g. -20 dBm) or too high (e.g. 30 dBm). We explain this phenomenon in detail as follows.
\par
Intuitively, when the BS transmit power is too low, the reflected signals by RIS are so weak that RISs have little contribution to performance improvement. While, when the BS transmit power is too high, the BSs prefer to allocate most of the power to the beams towards the BS-user links rather than the beams towards the BS-RIS-user links, which makes the role of \ac{ris}s less obvious. To be specific, as we can observe from Fig. \ref{img:simulation_3}, “{\it Without direct link}” can be roughly regarded as the scheme when the BSs allocate all power to the BS-RIS-user links, while “{\it Without RIS}” denotes the scheme when the BSs allocate all power to the BS-user links \cite{Wu'19}. With the increasing of BS transmit power, the performance gap between these two schemes becomes larger and larger. It indicates that, compared with the first scheme (i.e. BSs allocate all power to the BS-user links), the benefit of applying the second scheme (i.e. BSs allocate all power to the BS-RIS-user links) is relatively lower and lower. When the BS transmit power is large enough, the performance can obtain much more gain from BS allocating power to BS-user link than to BS-RIS-user link, since the performance gap between them is too large. In this case, the BS will tend to allocate most of its power to the beam towards the BS-user link, which weakens the role of RIS. 
\subsubsection{WSR against number of RIS elements}
\begin{figure}[!t]
	\centering
	\includegraphics[width=3.7in]{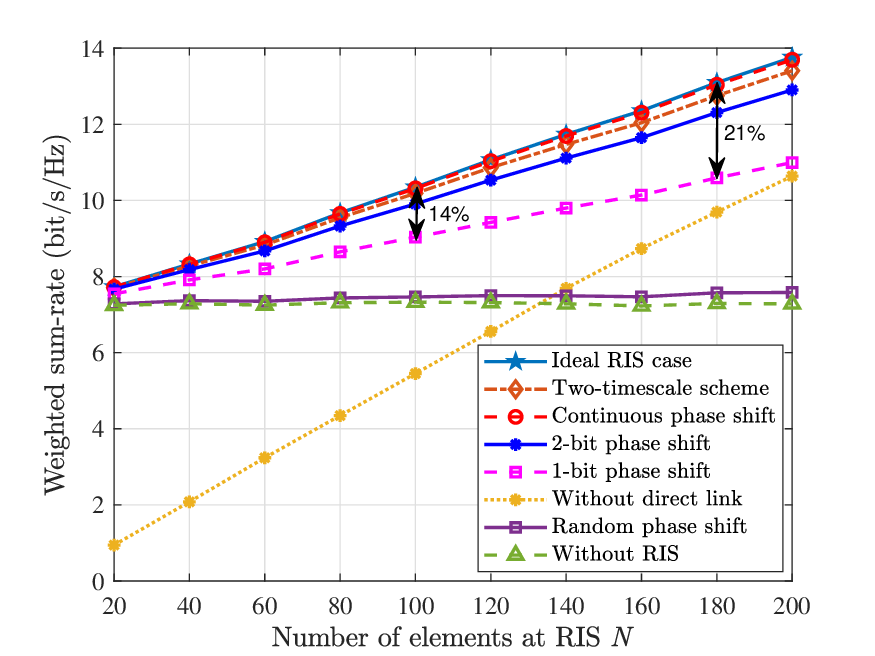}
	\caption{Weighted sum-rate against the number of RIS elements.}
	\label{img:simulation_4}
	\vspace{-0.8em}
\end{figure}
Using the same setups as above, we present the average weighted sum-rate against the number of RIS elements in Fig. \ref{img:simulation_4}. We can observe that, the \ac{wsr} of the proposed RIS-aided cell-free network increases as the number of \ac{ris} elements rises. More importantly, we find that, with the increasing of \ac{ris} elements, the approximation loss of the low-resolution phase shift becomes larger. For instance, when $N=100$, the approximation loss in the case “{\it 1-bit phase shift}” is about $14\%$ of the ideal case, and this loss grows to $21\%$ when $N=180$. The observation implies that, when the number of \ac{ris} is large, is is necessary to use more accurate phase shifts for passive precoding, so that the signals reflected from \ac{ris}s can reach the users more precisely. In addition, since the dimension of $\bf \Theta$ is $RN$, the complexity of solving (\ref{eqn:37}) grows with the number of \ac{ris} elements $N$, and too more RIS elements will also make the channel estimations more challenging. Thus, it is essential to choose the \ac{ris} element number reasonably.

\subsubsection{WSR with different allowable matched user-RIS pairs}
\begin{figure}[!t]
	\centering
	\includegraphics[width=3.7in]{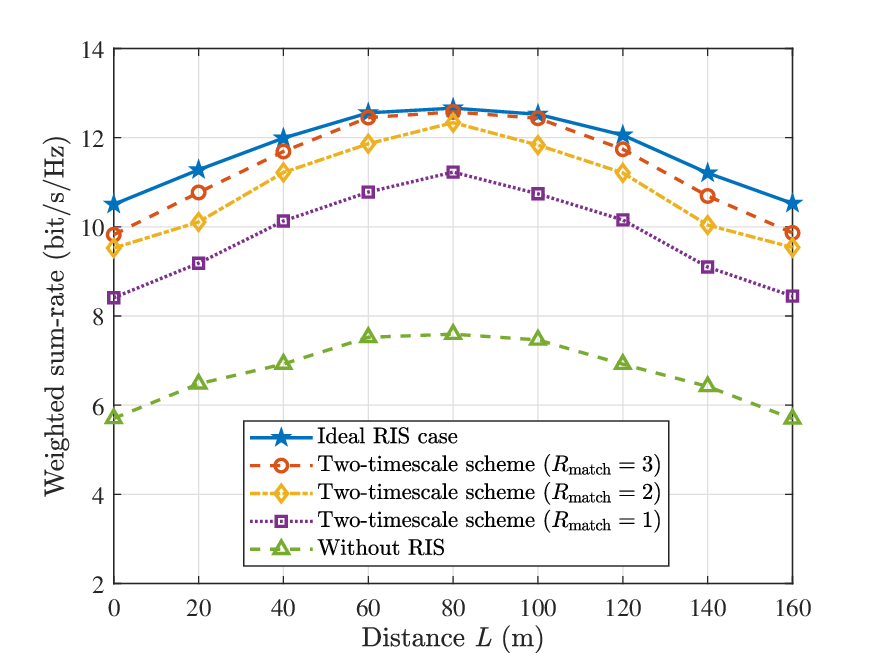}
	\caption{Weighted sum-rate against the distance $L$ with different allowable matched user-RIS pairs.}
	\label{img:simulation_two_time}
	\vspace{-0.8em}
\end{figure}

To evaluate the performance of the proposed two-timescale extension of the joint precoding scheme, in this part, we analyze the system WSR with different $R_{\rm match}$, which denotes the maximum number of RISs that each user can be matched with. To show more insights, in the same simulation scenario shown in Fig. \ref{img:simulation_scenario_1}, we reset the total number of RISs as $R=7$ and the RIS elements as $N=50$. We further assume that the $i$-th RIS is located at $\left(20\times i\,{\rm m},10\,{\rm m},6\,{\rm m}\right)$. Then, we plot the WSR against the distance $L$ with different $R_{\rm match}$ in Fig. \ref{img:simulation_two_time}. 

From this figure, we can observe that, the WSR increases as each user can be matched with more RISs. Compared with the ideal joint precoding design which acquires and utilizes all $KR=28$ RIS-aided channels, the proposed two-timescale scheme suffers performance but consumes much less overhead for CSI acquisitions. For example, when $R_{\rm match}=3$, i.e. each user can be matched with at most 3 RISs, compared with the perfect joint precoding design to some extent, the proposed two-timescale scheme acquires and utilizes no more than $KR_{\rm match}=12$ RIS-aided channels for joint precoding design but only suffers an averaged capacity loss of about $5\%$. Particularly, at the positions from $L=60\,{\rm m}$ to $L=100\,{\rm m}$, the two-timescale scheme ($R_{\rm match}=3$) can almost achieve the same performance as the ideal joint precoding design. In this way, the proposed two-timescale scheme consumes much less overhead for acquiring RIS-aided channels from the long-term perspective \cite{Huchen}. Thus, we can conclude that, the proposed two-timescale scheme can serve as a efficient scheme for balancing the performance and overhead for CSI acquisitions, which is especially suitable to employ in the cell-free networks with a large number of distributed RISs.

\subsection{Trade-off between the number of BSs and RISs}
	Since deploying more RISs can improve the network capacity with low cost and power consumption, in this subsection we analyze the trade-off between the number of BSs and RISs. 
	\par
	To fairly evaluate the losses and gains, here we take the energy efficiency as the performance metric. According to the power consumption model developed in \cite{Huang'18'2}, the system energy efficiency is given by
	\begin{equation}
		\begin{aligned}
			{E_{\rm sum}} = \frac{{{R_{\rm sum}}}}{\tau_{\rm r}{\left\| {\bf{W}} \right\|^2}+BP_{\rm BS} +KP_{\rm UE}
				+ RNP_{\rm RIS}},
		\end{aligned}
	\end{equation}
	where $\tau_{\rm r}^{-1}$ is the efficiency of the transmit power amplifier, while $P_{\rm BS}$, $P_{\rm UE}$, and $P_{\rm RIS}$ denote the hardware power consumption at each BS, each user, and each RIS element, respectively.
	\par
\begin{figure}[!t]
	\centering
	\includegraphics[width=3.7in]{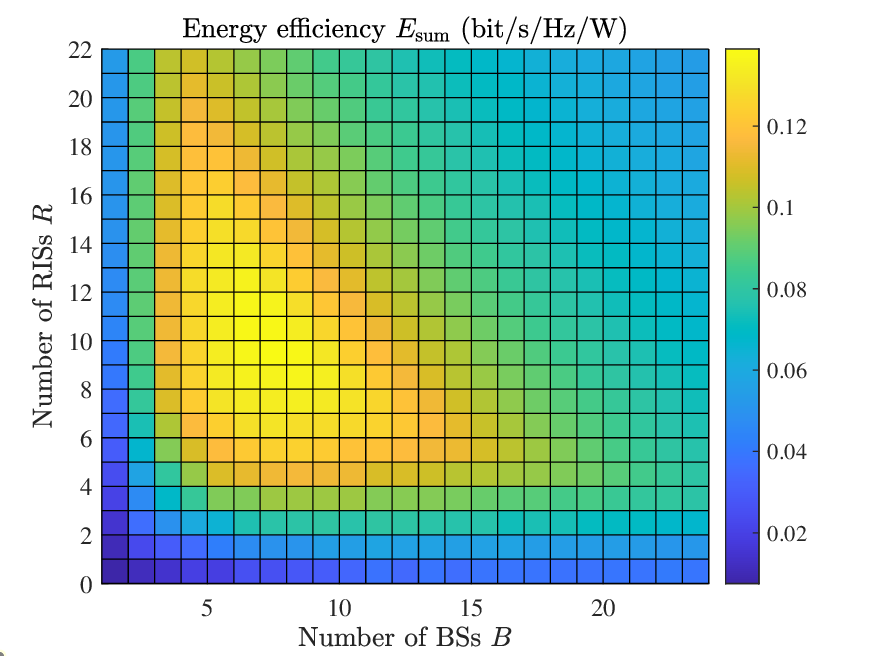}
	\caption{System energy efficiency against BS number $B$ and RIS number $R$.}
	\label{img:simulation_trade}
	\vspace{-0.8em}
\end{figure}

	For the simulation setups, we consider the same settings as those in \cite{Huang'18'2}, i.e. $\tau_{\rm r}=1.2$, $P_{\rm BS}=9$ dBW, $P_{\rm UE}=10$ dBm and $P_{\rm RIS}=10$ dBm. To find more essential insights, we reset some parameters as $P=M=U=1$ and $N=20$. We fix the distance-dependent effects by setting the related distances as $d_{\rm Bu}=d_{\rm BR}=110$ m and $d_{\rm Ru}=15$ m. By performing {\bf Algorithm 1} in ideal RIS case, we plot the energy efficiency against the number of BSs $B$ and that of RISs $R$ in Fig. \ref{img:simulation_trade}. From this figure, we can observe that, given $B$, the energy efficiency can be significantly improved by increasing the number of RISs $R$. Particularly, when $B=7$ and $R=9$, the maximal energy efficiency can be achieved, i.e. $E_{\rm sum}=0.138$ bit/s/Hz/W. However, when $R$ is too large, the energy efficiency will decline after deploying more RISs. For example, given $B=7$, the system energy efficiency decreases to  $E_{\rm sum}=0.09$ bit/s/Hz/W when $R=21$, which is about a 34.8\% loss. The reason is that, when the network capacity is large enough, the capacity improvement via deploying more RISs cannot make up for the extra power consumption. Therefore, carefully designing the trade-off between the number of BSs and RISs is the guarantee for balancing the capacity and power consumption in practical systems. 
\section{Conclusions and Future Works}\label{sec:con}
In this paper, we first propose the concept of RIS-aided cell-free network, which aims to improve the network capacity with low cost and power consumption. Then, for the proposed RIS-aided cell-free network, in a typical wideband scenario, we formulate the joint precoding design problem to maximize the weighted sum-rate to optimize the network capacity, subject to the transmit power constraint of \ac{bs} and the phase shift constraint of \ac{ris}. Finally, we propose a joint precoding framework to solve this problem. Since most of the considered scenarios in existing works are special cases of the general scenario in this paper, the proposed joint precoding framework can also serve as a general solution to maximize the capacity in most of existing RIS-aided scenarios. Simulation results demonstrate that, with the assist of low-cost and energy-efficient \ac{ris}s, the proposed RIS-aided cell-free network can realize higher capacity than the conventional cell-free network.
\par
Compared with the traditional cell-free network, the proposed RIS-aided cell-free network can achieve wider signal coverage \cite{Liang'19}, higher spectrum \cite{Wu'19}, and higher energy efficiency \cite{Huang'18'2}. For future works, some open problems are still worth further investigations. For example, some other important performance metrics such as the energy efficiency \cite{Huang'18'2}, BS transmit power \cite{Wu'19'2}, and user fairness \cite{Nadeem'19} are left for future works. Besides, some new hardware architecture such as active RIS \cite{Zhang'21} can be considered to further enhance the  capacity. In some specific scenarios such as the internet of vehicles, how to select the replaced BSs and optimize the placements of RISs is also an interesting research direction.

\par
Another promising technique benefiting from the advance of meta-materials is large intelligent surface (LIS) \cite{HuangHu'20,Dardari'20,Yuan'20}. Different from the passive RIS, LIS is considered as a quasi-continuous surface \cite{HuangHu'20} which can actively achieve surface beamforming \cite{Dardari'20}. Moreover, the signal model of LIS is in continuous-integral form \cite{HuangHu'20,Dardari'20,Yuan'20}. The introduction of LIS into cell-free network will bring many open problems, such as how to bridge the gap between discrete model and continuous model, which are also left for future works.

\appendices
	\section{PDS-based method for solving subproblem (\ref{eqn:25})}\label{appendix:1}
	We consider to solve (\ref{eqn:25}) via the PDS method \cite{Boyd'14}.
	Firstly, by introducing Lagrange multipliers ${\bm{\lambda }} = \left[ {{\lambda _1}, \cdots ,{\lambda _B}} \right]^T\in \mathbb{R}^{B}$ and $\vartheta>0$, the Lagrangian for the augmented problem of (\ref{eqn:25}) can be written as
	\begin{equation}
		\label{eqn:50}
		\begin{aligned}
			{\Gamma_1 ({\bf{W}},{\bm{\lambda }})}=g_{3}({\bf{W}})+{\bm \lambda}^T{\bf{F}}\left( {\bf{W}} \right)+\frac{\vartheta }{2}\left\| {{\bf{F}}\left( {\bf{W}} \right)} \right\|^2.
		\end{aligned}
	\end{equation}
	where the function vector ${\bf{F}}\left( {\bf{W}} \right)\in{\mathbb R}^B$ is defined by ${\bf{F}}\left( {\bf{W}} \right) = {\left[ {F_1^ + \left( {\bf{W}} \right), \cdots ,F_B^ + \left( {\bf{W}} \right)} \right]^T}$ with ${F_b}\left( {\bf{W}} \right)$ being ${F_b}\left( {\bf{W}} \right) = {{{{\bf{W}}^H}{{\bf{D}}_b}{\bf{W}} - {P_{b,\max }}} }$.
	\par
	Next, we introduce a superscript $t$ to each variable as the iteration
	index, and we have the iterative formulas of ${\bf W}^t$ and ${\bm \lambda}^t$ written as
	\begin{equation}\label{eqn:W_lambda_update}
		\begin{aligned}
			\left[ {\begin{array}{*{20}{c}}
					{{{\bf{W}}^{t + 1}}}\\
					{{{\bm \lambda} ^{t + 1}}}
			\end{array}} \right]{{ = }}\left[ {\begin{array}{*{20}{c}}
					{{{\bf{W}}^t}}\\
					{{{\bm \lambda} ^t}}
			\end{array}} \right]{{ - }}{\alpha ^t}\left[ {\begin{array}{*{20}{c}}
					{{{\left. {{{\partial {\Gamma _1}} \mathord{\left/
											{\vphantom {{\partial {\Gamma _1}} {\partial {\bf{W}}}}} \right.\kern-\nulldelimiterspace} {\partial {\bf{W}}}}} \right|}_{{\bf{W}} = {{\bf{W}}^t}}}}\\
					-{{{\left. {{{\partial {\Gamma _1}} \mathord{\left/
											{\vphantom {{\partial {\Gamma _1}} {\partial {\bf{\lambda }}}}} \right.
											\kern-\nulldelimiterspace} {\partial {\bm{\lambda }}}}} \right|}_{{\bm \lambda}  = {{\bm \lambda} ^t}}}}
			\end{array}} \right]
		\end{aligned}
	\end{equation}
	where the step length $\alpha^t>0$ is a sufficiently small positive value and the iteration directions ${{\partial {\Gamma _1}} \mathord{\left/
			{\vphantom {{\partial {\Gamma _1}} {\partial {\bf{W}}}}} \right.\kern-\nulldelimiterspace} {\partial {\bf{W}}}}$ and ${{\partial {\Gamma _1}} \mathord{\left/
			{\vphantom {{\partial {\Gamma _1}} {\partial {\bf{\lambda }}}}} \right.
			\kern-\nulldelimiterspace} {\partial {\bm{\lambda }}}}$ are given by
	\begin{subequations}
		\begin{align}
			\!\frac{{\partial {\Gamma _1}}}{{\partial {\bf{W}}}}&\!=\! {\left( {{\bf{AW}}} \right)^*} \!-\! {{\bf{V}}^*} \!+\! \sum\limits_{b = 1}^B \left( {{\lambda _b} \!+\! \vartheta F_b^ + \left( {\bf{W}} \right)} \right){{{\bf{\Pi }}_b^*}\left( {\bf{W}} \right)},\\
			\frac{{\partial {\Gamma _1}}}{{\partial {\bm{\lambda }}}} &\!=\! {\bf F}\left( {\bf{W}} \right),
		\end{align}
	\end{subequations}
	wherein the auxiliary function ${{\bf{\Pi }}_b}\left( {\bf{W}} \right)\in{\mathbb C}^{BMPK}$ is denoted by
	\begin{align}
		{{\bf{\Pi }}_b}\left( {\bf{W}} \right) = \left\{ {\begin{array}{*{20}{l}}
				{{{\bf{D}}_b}{\bf{W}},}&{{F_b}\left( {\bf{W}} \right) > 0}\\
				{{{\bf{0}}_{BMPK}},}&{{F_b}\left( {\bf{W}} \right) \le 0}
			\end{array},~~\forall b} \right. \in {\cal B}.
	\end{align}
	By simultaneously optimizing $\bf W$ and $\bm \lambda$ via (\ref{eqn:W_lambda_update}) until convergence, the optimal ${\bf W}^{\rm opt}$ can be finally obtained without need of the inversion operation for the high-dimensional matrix $\bf A$. 
	\begin{remark}\label{Remark:appendix2}
		We discuss the computational complexity of solving (\ref{eqn:25}). Since ${\bf D}_b$ is a fixed diagonal matrix with $MPK$ elements being 1, we can simply derive that the complexity of computing ${F_b^+}\left( {\bf{W}} \right)$ is ${\cal O}\left(MPK\right)$. Then, we can further obtain that the computational complexity of updating $\bf W$ and $\bm \lambda$ via (\ref{eqn:W_lambda_update}) are ${\cal O}\left(B^2M^2P^2K^2\right)$ and ${\cal O}\left(BMPK\right)$, respectively. Let $I_{\rm a}$ denote the required iteration number for convergence. Then the overall complexity of the applied PDS-based method is about ${\cal O}\left( {{I_{\rm{a}}}\left( {{B^2}{M^2}{P^2}{K^2} + BMPK} \right)} \right)$.		
	\end{remark}	
	
	\section{PDS-based method for solving subproblem (\ref{eqn:37})}\label{appendix:2}
	We consider to solve (\ref{eqn:37}) via the PDS method \cite{Boyd'14}.
	Similar to Appendix \ref{appendix:1}, by introducing Lagrange multipliers ${\bm{\chi }} = {\left[ {{\chi _{1}}, \cdots ,{\chi _{RN}}} \right]^T} \in {\mathbb{R}^{RN}}$ and $\varphi>0$, the Lagrangian for the augmented problem of (\ref{eqn:37}) can be derived as
	\begin{equation}
		\begin{aligned}
			{\Gamma_2 ({\bf{\Theta}},{\bm{\chi }})}=g_{6}({\bf{\Theta}})+{\bm \chi}^T{\bf{G}}\left( {\bf{\Theta}} \right)+\frac{\varphi }{2}\left\| {{\bf{G}}\left( {\bf{\Theta}} \right)} \right\|^2,
		\end{aligned}
	\end{equation}
	where the function vector ${\bf{G}}\left( {\bf{\Theta}} \right)$ is defined by ${\bf{G}}\left( {\bf{\Theta}} \right) = {\left[ {G_1^ + \left( {\bf{\Theta}} \right), \cdots ,G_{RN}^ + \left( {\bf{\Theta}} \right)} \right]^T}$ with ${G_j}\left( {\bf{\Theta}} \right) = {{\bm{ \theta} }^H}{\bf E}_j{\bm{ \theta} }-1$ and ${\bf E}_j\triangleq{\bf{e}}_{j}{\bf{e}}_j^{H}$. Then, we have the iterative formulas of ${\bm \theta}^t$ and ${\bm \chi}^t$ written as
	\begin{equation}\label{eqn:Theta_chi_update}
		\begin{aligned}
			\left[ {\begin{array}{*{20}{c}}
					{{{\bm{\theta}}^{t + 1}}}\\
					{{{\bm \chi} ^{t + 1}}}
			\end{array}} \right]{{ = }}\left[ {\begin{array}{*{20}{c}}
					{{{\bm{\theta}}^t}}\\
					{{{\bm \chi} ^t}}
			\end{array}} \right]{{ - }}{\beta ^t}\left[ {\begin{array}{*{20}{c}}
					{{{\left. {{{\partial {\Gamma _2}} \mathord{\left/
											{\vphantom {{\partial {\Gamma _1}} {\partial {\bm{\theta}}}}} \right.\kern-\nulldelimiterspace} {\partial {\bm{\theta}}}}} \right|}_{{\bm{\theta}} = {{\bm{\theta}}^t}}}}\\
					-{{{\left. {{{\partial {\Gamma _2}} \mathord{\left/
											{\vphantom {{\partial {\Gamma _1}} {\partial {\bf{\chi }}}}} \right.
											\kern-\nulldelimiterspace} {\partial {\bm{\chi }}}}} \right|}_{{\bm \chi}  = {{\bm \chi} ^t}}}}
			\end{array}} \right]
		\end{aligned}
	\end{equation}
	wherein $\beta^t>0$ denotes the sufficiently small step length and the iteration directions ${{\partial {\Gamma _2}} \mathord{\left/
			{\vphantom {{\partial {\Gamma _2}} {\partial {\bm{\theta}}}}} \right.\kern-\nulldelimiterspace} {\partial {\bm{\theta}}}}$ and ${{\partial {\Gamma _2}} \mathord{\left/
			{\vphantom {{\partial {\Gamma _2}} {\partial {\bm{\chi }}}}} \right.
			\kern-\nulldelimiterspace} {\partial {\bm{\chi }}}}$ are obtained by
	\begin{subequations}
		\begin{align}
			\!\frac{{\partial {\Gamma _2}}}{{\partial {\bm{\theta}}}}&\!=\! {\left( {{\bm{\Lambda}\bm{\theta}}} \right)^*} - {{\bm{\nu}}^*} + \sum\limits_{j = 1}^{RN}  \left( {{\chi _j} + \varphi G_j^ + \left( {\bm{\Theta}} \right)} \right){{{\bf{\Psi }}_j^*}\left( {\bm{\theta }} \right)},\\
			\frac{{\partial {\Gamma _2}}}{{\partial {\bm{\chi }}}} &\!=\! {\bf G}\left( {\bm{\Theta}} \right).
		\end{align}
	\end{subequations}
	Note that, the auxiliary function ${{\bf{\Psi }}_j}\left( {\bm{\theta }} \right)\in{\mathbb C}^{RN}$ is defined as
	\begin{align}
		{{\bf{\Psi }}_j}\left( {\bm{\theta }} \right) \!=\! \left\{ {\begin{array}{*{20}{l}}
				{{{\bf{E}}_j}{\bm{\theta }},}&{{{\bm{\theta }}^H}{{\bf{E}}_j}{\bm{\theta }} > 0}\\
				{{{\bf{0}}_{RN}},}&{{{\bm{\theta }}^H}{{\bf{E}}_j}{\bm{\theta }} \le 0}
		\end{array}} \right.,\forall j\in \left\{ {1, \cdots ,RN} \right\}
	\end{align}
	By simultaneously optimizing $\bm \theta$ and $\bm \chi$ via (\ref{eqn:Theta_chi_update}) until convergence, the optimal ${\bm \Theta}^{\rm opt}$ can be finally obtained without need of the inversion operation for the high-dimensional matrix $\bf {\bm \Lambda}$.
	\begin{remark}\label{Remark:appendix3}
		We study the computational complexity of solving (\ref{eqn:37}). Similar to {\bf Remark \ref{Remark:appendix2}}, we can easily derive that the complexity of updating $\bm \Theta$ and $\bm \chi$ by (\ref{eqn:Theta_chi_update}) are ${\cal O}\left(R^2N^2\right)$ and ${\cal O}\left(RN\right)$, respectively. Therefore, the overall computational complexity is ${\cal O}\left( {{I_{\rm{p}}}\left( {{R^2}{N^2} + RN} \right)} \right)$ where $I_{\rm p}$ denotes the required iteration number for convergence.
	\end{remark}	
\footnotesize
\bibliographystyle{IEEEtran}
\bibliography{IEEEabrv,reference}
\input{biographies}
\end{document}

%% file: biographies.tex
\begin{IEEEbiography}[{\includegraphics[width=1in,height=1.25in,clip,keepaspectratio]{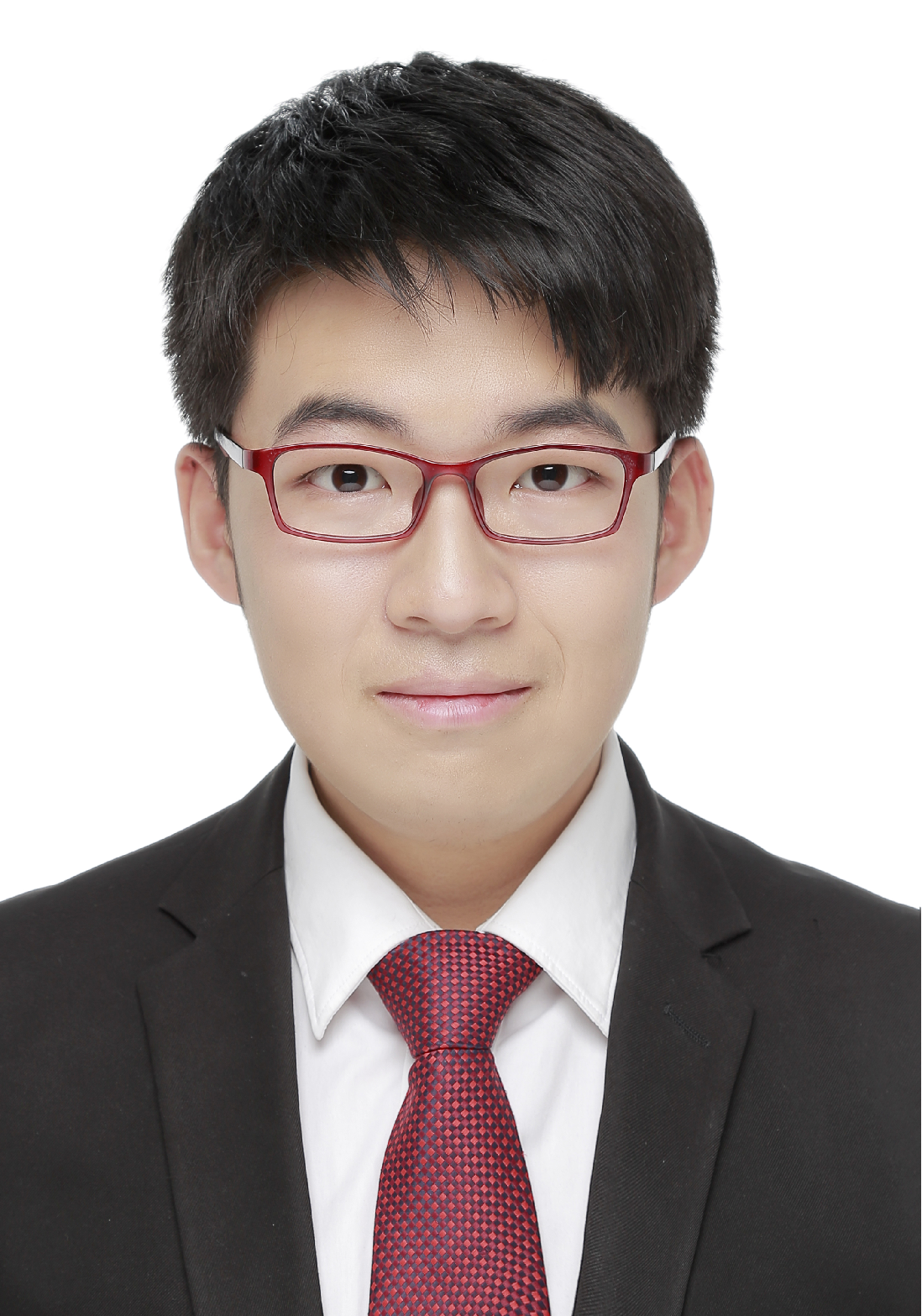}}]{Zijian Zhang}
(S'20) received the B.E. degree in electronic engineering from Tsinghua University, Beijing, China, in 2020. He is currently working toward the Ph.D. degree in electronic engineering from Tsinghua University, Beijing, China.
His research interests include physical-layer algorithms for massive MIMO and reconfigurable intelligent surfaces (RIS). He has received the National Scholarship in 2019 and the Excellent Thesis Award of Tsinghua University in 2020.
\end{IEEEbiography}

\begin{IEEEbiography}[{\includegraphics[width=1in,height=1.25in,clip,keepaspectratio]{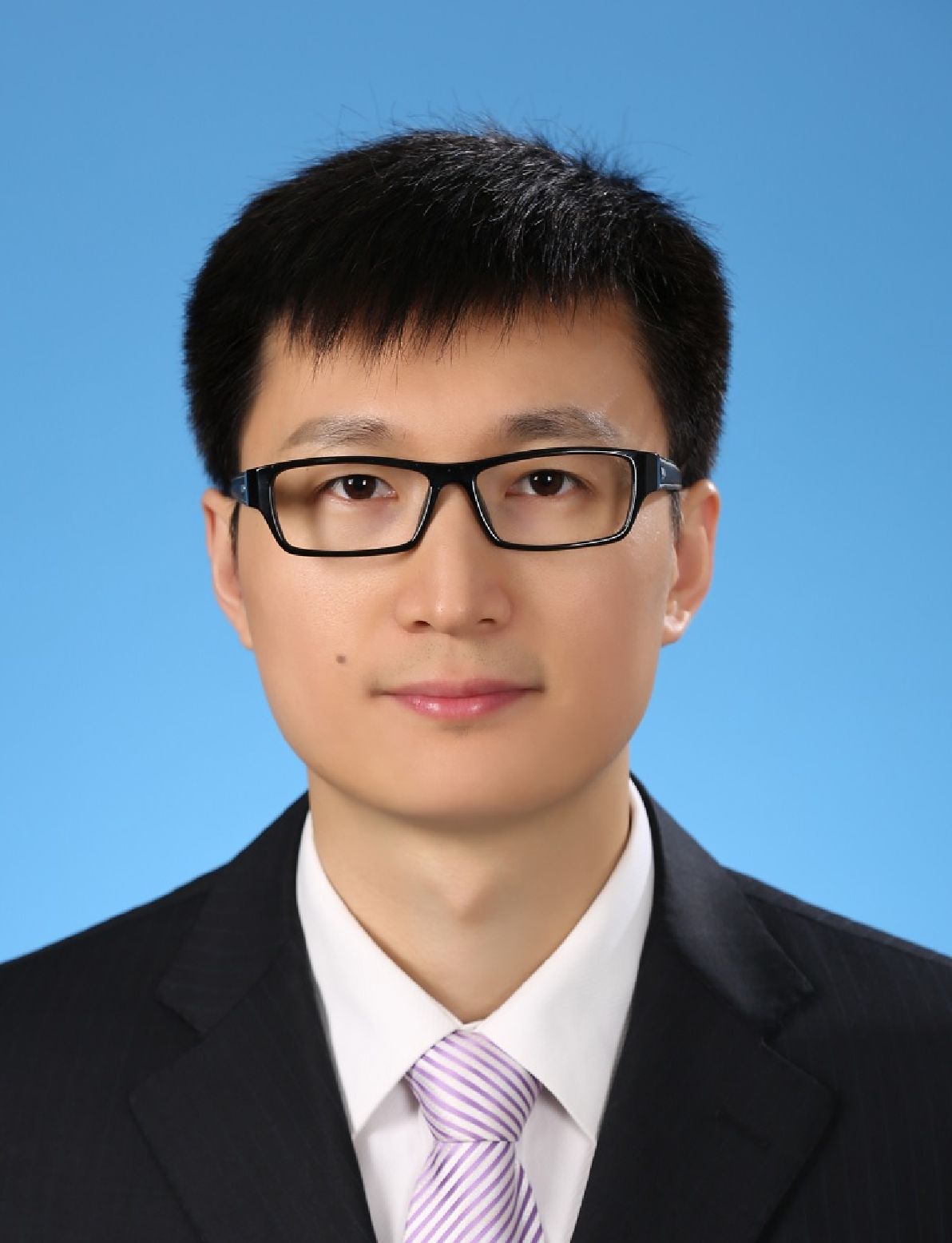}}]{Linglong Dai} (M'11-SM'14) received the B.S. degree from Zhejiang University, Hangzhou, China, in 2003, the M.S. degree (with the highest honor) from the China Academy of Telecommunications Technology, Beijing, China, in 2006, and the Ph.D. degree (with the highest honor) from Tsinghua University, Beijing, China, in 2011. From 2011 to 2013, he was a Postdoctoral Research Fellow with the Department of Electronic Engineering, Tsinghua University, where he was an Assistant Professor from 2013 to 2016 and has been an Associate Professor since 2016. His current research interests include reconfigurable intelligent surface (RIS), massive MIMO, millimeter-wave/ Terahertz communications, and machine learning for wireless communications.

He has coauthored the book ``MmWave Massive MIMO: A Paradigm for 5G" (Academic Press, 2016). He has authored or coauthored over 60 IEEE journal papers and over 40 IEEE conference papers. He also holds 19 granted patents. He has received five IEEE Best Paper Awards at the IEEE ICC 2013, the IEEE ICC 2014, the IEEE ICC 2017, the IEEE VTC 2017-Fall, and the IEEE ICC 2018. He has also received the Tsinghua University Outstanding Ph.D. Graduate Award in 2011, the Beijing Excellent Doctoral Dissertation Award in 2012, the China National Excellent Doctoral Dissertation Nomination Award in 2013, the URSI Young Scientist Award in 2014, the IEEE Transactions on Broadcasting Best Paper Award in 2015, the Electronics Letters Best Paper Award in 2016, the National Natural Science Foundation of China for Outstanding Young Scholars in 2017, the IEEE ComSoc Asia-Pacific Outstanding Young Researcher Award in 2017, the IEEE ComSoc Asia-Pacific Outstanding Paper Award in 2018, the China Communications Best Paper Award in 2019, the IEEE Access Best Multimedia Award in 2020, and the IEEE Communications Society Leonard G. Abraham Prize in 2020. He was listed as a Highly Cited Researcher by Clarivate Analytics in 2020.

He is an Area Editor of IEEE Communications Letters, and an Editor of IEEE Transactions on Communications and IEEE Transactions on Vehicular Technology. Particularly, he is dedicated to reproducible research and has made a large amount of simulation codes publicly available.

\end{IEEEbiography}